\input phyzzx
\hsize=40pc
\font\sf=cmss10                    

\def\Figure#1#2{\midinsert
$$\BoxedEPSF{#1}$$
\noindent {\sf #2}
\endinsert}

\input BoxedEPS
\SetRokickiEPSFSpecial 
\HideDisplacementBoxes
\catcode`\@=11 
\def\NEWrefmark#1{\step@ver{{\;#1}}}
\catcode`\@=12 

\def\square{\kern1pt\vbox{\hrule height 1.2pt\hbox{\vrule width 1.2pt\hskip 3pt
   \vbox{\vskip 6pt}\hskip 3pt\vrule width 0.6pt}\hrule height 0.6pt}\kern1pt}

\def\bra#1{\langle #1 |}
\def\ket#1{| #1 \rangle}

\def\ov{{\overline}}

\def\A{{\cal A}}
\def\B{{\cal B}}
\def\C{{\cal C}}
\def\D{{\cal D}}
\def\H{\widehat{\cal H}}

\def\I{{\cal I}}
\def\K{{\cal K}}

\def\M{{\cal M}}

\def\O{{\cal O}}
\def\P{{\cal P}}

\def\T{{\cal T}}

\def\V{{\cal V}}

\def\p{\partial}

\def\k{{\kappa}}

\def\ov{\overline}

\def\wh{\widehat}

\def\B{{\cal B}}

\def\P{{\cal P}}
\def\V{{\cal V}}
\def\O{{\cal O}}

\def\p{\partial}

\singlespace

\def\define#1#2\par{\def#1{\Ref#1{#2}\edef#1{\noexpand\refmark{#1}}}}
\def\con#1#2\noc{\let\?=\Ref\let\<=\refmark\let\Ref=\REFS
         \let\refmark=\undefined#1\let\Ref=\REFSCON#2
         \let\Ref=\?\let\refmark=\<\refsend}

\let\refmark=\NEWrefmark

\define\zwiebachlong{B. Zwiebach, `Closed string field theory: Quantum
action and the Batalin-Vilkovisky master equation', Nucl. Phys {\bf B390}
(1993) 33, hep-th/9206084.}

\define\senzwiebach{A.~Sen and B.~Zwiebach, `Local background
independence of classical closed string field theory', {\it Nucl.\
Phys.\ } {\bf B414} (1994) 649, hep-th/9307088.}

\define\senzwiebachtwo{A.~Sen and B.~Zwiebach, `Quantum background
independence of closed string field theory', {\it Nucl.\ Phys.\ } {\bf
B423} (1994) 580, hep-th/9311009.}

\define\senzwiebachgauge{A.~Sen and B.~Zwiebach, `A note on gauge
transformations in Batalin-Vilkovisky theory', {\it Phys.\ Lett.\ }
{\bf B320} (1994) 29, hep-th/9309027.}

\define\senzwiebachnew{A.~Sen and B.~Zwiebach, `Background
independent algebraic structures in closed string field theory',
MIT-CTP-2346, August 1994, hep-th/9408053.}

\define\zwiebachncon{B.~Zwiebach, `On the construction of string field theory
around non-conformal backgrounds', MIT-CTP-2531, to appear.}

\define\campbell{M.~Campbell, P.~Nelson and E.~Wong, `Stress tensor
perturbations in conformal field theory', {\it Int.\ Jour.\ Mod.\
Phys.\ } {\bf A6} (1991) 4909.}

\define\ranganathan{K.~Ranganathan, `Nearby CFT's in the operator
formalism: the role of a connection', {\it Nucl.\ Phys.\ } {\bf
B408} (1993) 180. }

\define\rangasonodazw{K.~Ranganathan, H.~Sonoda and B.~Zwiebach, `Connections
on the state-space over conformal field theories', {\it Nucl.\ Phys.\ }
{\bf B414} (1994) 405, hep-th/9304053.}

\define\gussichsundell{A. Gussich, and P. Sundell, `Finite deformations
of conformal field theories using analytically regularized connections',
G\"oteborg preprint, ITP 96-5, hep-th/9604010.}

\define\bergmanzwiebach{ O.Bergman and B.Zwiebach `The dilaton theorem
and closed string backgrounds' Nucl.~Phys. {\bf B441} (1995) 76,
hep-th/9411047.}

\define\getzler{E. Getzler, `Batalin-Vilkovisky algebras and two-dimensional
topological field theories', Commun. Math. Phys. {\bf 159} (1994) 265;
 hep-th/9212043.}

\define\lianzuckerman{B. H. Lian and G. Zuckerman,
``New Perspectives on the BRST-algebraic structure of string theory'',
 Commun. Math. Phys. {\bf 154} (1993) 613,  hep-th/9211072.}

\define\penkavaschwarz{M. Penkava and A. Schwarz,
``On Some Algebraic Structures Arising in String Theory",
UC Davis preprint, UCD-92-03, hep-th/9212072.}

\define\kimurastasheff{T. Kimura, J. Stasheff, and A. A. Voronov, `On operad
structures
of moduli spaces and string theory',  Commun. Math. Phys. {\bf 171} (1995) 1,
hep-th/9307114;\hfill\break
`Homology of moduli spaces of curves and commutative homotopy 
algebras' , alg-geom/9502006.}

\define\kimuravoronov{T. Kimura and A. A. Voronov, `The cohomology of algebras
over moduli spaces', University of North Carolina preprint,
October 1994, hep-th/9410108.}

\define\bering{K. Bering, P.H. Damgaard, and J. Alfaro, `Algebra of higher
antibrackets' hep-th/9604027.}

\define\akman{F. Akman, `On some generalizations of Batalin-Vilkovisky
algebras', q-alg/9506027}

\define\sonoda{H. Sonoda, Composite operators in QCD,
Nucl. Phys. {\bf B383} (1992) 173, hep-th/9205085;\hfill\break
``Operator Coefficients for Composite Operators in the $(\phi^4)_4$
Theory'', Nucl. Phys. {\bf B394} (1993) 302, hep-th/9205084. }

\define\kugozwiebach{T. Kugo and B. Zwiebach, 
Prog. Theo. Phys. {\bf 87} (1992) 801}

\define\cornalba{L. Cornalba, Connections over spaces of conformal field
theories: testing for integrability, M.S. Thesis, MIT, May 1994.}

\define\barnich{G. Barnich, F. Brandt, and M. Henneaux, `Local BRST cohomology
in Einstein-Yang-Mills theory', hep-th/9505173;
`Local BRST cohomology in the antifield formalism: I. General theorems',
hep-th/9405109; `Local BRST cohomology in the antifield formalism: II
Application to Yang-Mills theory', hep-th/9405194.}

\singlespace
{}~ \hfill \vbox{\hbox{MIT-CTP-2527}
\hbox{
} }\break
\title{ NEW  MODULI SPACES FROM  STRING BACKGROUND}
\titlestyle{ INDEPENDENCE  CONSISTENCY CONDITIONS}
\author{Barton Zwiebach \foot{E-mail address: zwiebach@irene.mit.edu
\hfill\break Supported in part by D.O.E.
cooperative agreement DE-FC02-94ER40818.}}
\address{Center for Theoretical Physics,\break
Laboratory of Nuclear Science\break
and Department of Physics\break
Massachusetts Institute of Technology\break
Cambridge, Massachusetts 02139, U.S.A.}

\abstract 
{In string field theory an infinitesimal
background deformation is implemented as a canonical 
transformation whose hamiltonian function is defined by  
moduli spaces of punctured Riemann surfaces  having one special puncture.
We show that the consistency conditions associated to the commutator
of two deformations are implemented by
virtue of the existence of moduli spaces of punctured surfaces 
with two special punctures. The spaces are antisymmetric
under the exchange of the special punctures, and satisfy recursion relations
relating them to moduli spaces with one special puncture and 
to string vertices. We develop the theory of moduli spaces of surfaces 
with  arbitrary number of special punctures and indicate their 
relevance to the construction of a string field theory that makes 
no reference to a conformal background. Our results also imply a partial
antibracket cohomology theorem for the string action.}
\endpage

\singlespace
\baselineskip=18pt

\chapter{Introduction and Summary}

Finding the conceptual framework for string theory is an outstanding
open problem of great relevance. One way to search for this
framework is to develop and extend as far as is possible the 
present formulation of string field theory. 
This present formulation uses explicitly a chosen conformal background
and is  not manifestly independent of this choice. 
Nevertheless this admittedly incomplete string field theory has revealed a
rather elegant structure. The mathematical framework of 
the present-day field theory of strings is by now fairly clear. It contains
an ingredient from Riemann surfaces and an ingredient from conformal theory,
with similar and related structures existing in both.

Consider first the Riemann surface ingredient.
For moduli spaces of punctured surfaces, 
the natural operation of cutting 
disks around  punctures and sewing the resulting boundaries with a twist allow 
one to define the antibracket $\{ \, , \, \}$ of two moduli spaces of surfaces 
and a delta operation $\Delta$  on a moduli space of surfaces. 
These operations
satisfy the axioms of the Batalin-Vilkovisky (BV) antibracket and 
delta operation.   
Moreover,  the sets of moduli spaces that define string vertices can be
grouped into an element called $\V$, and the
condition that the corresponding Feynman graphs generate a cover of all
moduli spaces of Riemann surfaces implies that $\V$ can be used to define
a cohomology class for the operator $\partial + \Delta$, where $\partial$
is the boundary operator on moduli spaces of surfaces. 

Consider now the conformal theory ingredient. A suitable
conformal field 
theory furnishes a vector space $\H$ spanned by  all local operators,
and equipped with an antibracket and
a delta operation defined on the functions on $\H$. Moreover it furnishes
an odd quadratic hamiltonian $Q$ satisfying $\{ Q , Q \}=0$ arising from the
BRST operator on $\H$. There is, moreover, a map $f$ from moduli spaces of
surfaces to functions on $\H$. This map induces a homomorphism between
the BV structure on moduli spaces and the BV structure on $\H$, with
the boundary operator $\partial$ becoming the BRST hamiltonian $Q$. Finally
the string action $S$ is simply given as $S = Q + f(\V)$, and
is a function that manifestly satisfies the BV master equation, thus
defining a gauge invariant action that can be quantized consistently. 

The above viewpoint on string field theory was explained precisely in 
Ref.[\senzwiebachnew] and was obtained while investigating the problem of
background independence [\senzwiebach,\senzwiebachtwo,\senzwiebachgauge].
There are certainly other ways of viewing string field theory, where one
emphasizes other aspects of the algebraic structure or other
aspects of the geometrical foundation of the theory 
[\kimurastasheff,\kimuravoronov,\zwiebachlong]. 
For related matters see also
Refs.[\getzler,\lianzuckerman,\penkavaschwarz], and for recent discussion see
Refs.[\bering,\akman].

Since present-day string field theory is not manifestly 
independent of the choice of background needed for its formulation, the above
structure cannot be expected to be the final word. In fact, the analysis of
background independence of  Refs.[\senzwiebach,\senzwiebachtwo] brought into
the open an operator
$\K$ which acting on a Riemann surface adds one special puncture throughout
the surface minus the unit disks around the punctures (the disks that define
the local coordinates around the punctures). This operation is intimately
related, at the level of spaces of conformal theories,
to the operation of covariant differentiation using a particular 
connection [\campbell,\ranganathan,\rangasonodazw,\sonoda,\gussichsundell].
In addition to the operator $\K$, a new family of moduli spaces of Riemann
surfaces was introduced, spaces where the surfaces have one special puncture
in addition to the ordinary punctures. These spaces, called $\B^1$ spaces
(our earlier notation used no superscript), have
a real dimensionality that exceeds by one that of the corresponding
moduli space with only
ordinary punctures. More precisely,  the moduli space $\B_{g,n-1}^1$ having
$(n-1)$ ordinary punctures and one special one, has dimension equal to
that of $\M_{g,n}$ plus one (that is $6g-6+2n+1$) . Roughly speaking the
$\B_{g,n-1}^1$ space interpolates between the string vertex $\V_{g,n}$ where
one of the punctures is considered special, and $\K\V_{g,n-1}$ which is
the ``vertex" obtained by adding one special puncture via $\K$ to a vertex
with one less puncture. 

The $\B^1$ spaces were the Riemann surface ingredient
for the proof of background independence of string field theory. Inserting
a marginal operator  $\wh\O_\mu$ in the special puncture, the $\B^1$ spaces 
define
a hamiltonian function $B_\mu$ (formerly called $U_\mu$) that by canonical
action via the antibracket generates a change of string background. Given that
they play such prominent role in the proof of local background independence it
was suggested in  Ref.[\senzwiebachtwo] that $\B^1$ spaces could also play a
role in the  construction that makes background independence manifest.
 
Here we indicate a natural way the suggestion can be
realized if we attempt to construct a string field theory around a
non-conformal background.  A string field theory written  
using a non-conformal
background would represent concrete progress
since the  main problem with the present formulation is not that we 
use a background to write the
theory, but rather the fact that the background needs to be conformal.
The algebraic structure of string theory formulated around non-conformal
backgrounds  was explored by indirect means in 
Ref.[\zwiebachlong] sect 4.5. This structure requires a special
state $F$ of ghost number three whose vanishing would imply that we
are again considering a conformal theory. 
This special string field $F$ modifies
significantly the identities that must be satisfied by the string
vertices. We now recall that ordinary closed string fields are of ghost number
two and ghost number conservation requires correlations of
ordinary closed string fields 
to be integrated on moduli spaces of standard dimensionality. Being
of ghost number three in order to couple $F$ to ordinary string fields 
we require moduli spaces of real dimension one higher than the standard
dimension. As mentioned in the previous paragraph $\B^1$ spaces have 
precisely this property! The way to use the $\B^1$ spaces to construct
explicitly an action representing string theory around non-conformal 
backgrounds will be the subject of a forthcoming 
publication [\zwiebachncon].

It can be guessed, however, that the presently known 
$\B^1$ spaces cannot be the complete story.
Moduli spaces whose surfaces have only one special puncture clearly
represent first order perturbations of the background. 
We should expect that moduli spaces $\B^2,\B^3,\cdots $ of surfaces with
two, with three, in fact, with all numbers of special punctures 
should be relevant [\zwiebachncon].
We must therefore learn how to deal with moduli spaces whose surfaces 
contain more than one special puncture, and this is the main goal of the
present paper. The case of two special punctures brings in most of the
new features of the problem, and therefore much of our work here will
have to do with this case.  Luckily, this case need not be
investigated in  the difficult context of constructing a more general string
field theory.

In this paper we show that moduli spaces $\B^2$ of surfaces with
two special punctures make an appearance when we consider consistency
conditions arising from the {\it commutator} of two infinitesimal background
deformations. 
Arising from a commutator, $\B^2$ spaces should be antisymmetric
under the exchange of the two special punctures. This is also in accord with
our expectation for such spaces in a new formulation of string field theory;
if we are to insert ghost number three grassmann odd $F$ states on every 
special puncture, moduli spaces of surfaces with special punctures  
must be odd under the exchange of any pair of special punctures.\foot{In 
the analysis of background independence of Refs.[\senzwiebach,\senzwiebachtwo] 
one inserted a grassman even marginal
state on the special puncture of the $\B$ spaces. The construction of
hamiltonians for large background deformations may require
moduli spaces symmetric under the exchange of the special punctures.}
This antisymmetry brings about  a number of simplifications. 

When we examine the consistency conditions for background independence
we obtain a condition of the form $\{ S, H_{\mu\nu}\} =0$, 
where $S$ is
the string action. $H_{\mu\nu}$ is an object built from
covariant derivatives of the hamiltonian $B_\mu$ 
generating a
background deformation in the direction of the marginal
operator $\wh\O_\mu$, and from the curvature $R_{\mu\nu}$  of a
theory  space connection. While it is clear from our arguments that the
consistency condition is satisfied, we demand a stronger result, we 
require that $H_{\mu\nu} = \{ S, B_{\mu\nu}\}$, 
where $B_{\mu\nu}$ is
a hamiltonian to be determined. It then follows that
$\{ S , \{ S , B_{\mu\nu}\} \} =0$ by
virtue of the Jacobi identity and the master equation $\{S , S\} =0$. The
existence of $B_{\mu\nu}$
amounts to a cohomology theorem for the string action $S$, in this case
saying that the $S$-closed function $H_{\mu\nu}$ is $S$ exact. 
Our goal is not only to show that the hamiltonian $B_{\mu\nu}$ exists,
but also to show that it is simply a hamiltonian
associated to moduli spaces $\B^2$ of surfaces having two special punctures.
This is certainly not manifest since the object $H_{\mu\nu}$ appears
to involve the theory space connection in a nontrivial way. 

Our results, apart from their significance on the context of
building a more general string field theory,  represent some
progress towards developing a complete understanding of the antibracket
cohomology defined by the string action $S$. In
fact, even the existence (shown in [\senzwiebach]) of the hamiltonian
$B_\mu$ generating the infinitesimal string background deformations can
be viewed as a cohomology theorem.
Cohomology theorems are useful to understand issues of 
uniqueness, anomalies,
and renormalization. Complete antibracket cohomology theorems are available for
theories of nonabelian gauge fields and gravitation [\barnich].

In order to understand the moduli space origin of the hamiltonian
$B_{\mu\nu}$ we must develop 
a little the geometrical understanding of
connections achieved in Ref.[\rangasonodazw]. In particular we show that the
reference  connection
$\Gamma$ (called  $\wh\Gamma$ in [\rangasonodazw]) is such that 
the antisymmetric combination of covariant derivatives
$D_\mu \ket{\wh\O_\nu} -D_\nu
\ket{\wh\O_\mu}$ is simply related to a moduli space $\T^2_1$ of spheres with
 one  ordinary puncture and two special punctures
having the states
$\wh\O_\mu$ and $\wh\O_\nu$  inserted in them. Moreover, we give a simple
proof that  the connection $\Gamma$ has zero curvature.  

We generalize our earlier definition of the antibracket of
two moduli spaces of surfaces to include the case when the moduli spaces
are comprised of surfaces
with special punctures. We introduce a new operator $\I$ whose effect
is to turn an ordinary puncture into a special puncture preserving the
antisymmetry of all the special punctures. Since moduli
spaces are symmetric under the exchange of ordinary punctures $\I^2=0$
follows because repeated action by $\I$ will turn two ordinary punctures
special. We also find a suitable definition for $\K$ acting on surfaces having
special punctures. The operator $\K$ will insert a new puncture on the
surface minus the unit disks around the {\it ordinary} punctures; it
ignores the presence of the special punctures. The collisions between
the old special punctures and the moving special puncture are rendered
harmless by the antisymmetry condition. Moreover we show that $\K^2=0$,
a property that is intimately related  the zero curvature property
of the connection $\Gamma$. We develop a series of identities
relating the operators $\K$, $\I$ and the antibracket.    

Equipped with the above tools we can show that the hamiltonian 
$B_{\mu\nu}$
is indeed defined by moduli spaces of surfaces with two special punctures,
moduli spaces denoted collectively as $\B^2 = \sum_{n\geq 1}
\B^2_n~$, where
$\B^2_n$ is a moduli space of spheres with two special punctures
and $n$ ordinary  punctures. Similarly, we write
$\B^1=\sum_{n\geq 2} \B^1_n$ where $\B^1_n$ is a moduli space of 
spheres with one special puncture
and $n$ ordinary  punctures. The 
equations that constrain $\B^2$ take the form of recursion relations 
involving the $\B^1$ spaces and the string vertices $\V$. They  read
$$  \partial\B^2  = \T^2_1 +  (\K - \I )  \,\B^1  -\, \half\,
 \{ \B^1 \,,\, \B^1 \} - \{ \V\, , \,\B^2 \}\,. \eqn\ghyto$$
We prove the consistency of the above equations by showing that the
right hand side is annihilated by the boundary operator $\partial$. We then
give a recursive construction of the $\B^2$ spaces.

All our work will concern only the classical string action $S$. No
particular difficulties are expected for the extension to the full quantum
master action, apart from the usual subtleties associated to vacuum vertices,
particularly at genus one. While we give here the general framework to
deal with moduli spaces having any number of special punctures, the
explicit conditions defining useful moduli spaces $\B^3, \B^4, \cdots$, will
be given in Ref.[\zwiebachncon]. As the analysis of the present work
suggests, they can be obtained as higher cohomology theorems resulting
from multiple commutators of background deformations. 

This paper is organized as follows. In sect.2 we review some necessary
results on connections on spaces of conformal field theories and covariant
derivatives. We also recall the main properties of the string vertices $\V$ and
the $\B^1$ spaces. In sect.3 we derive and examine  the consistency conditions
arising from the commutator of two infinitesimal background deformations.
In sect.4 we develop further our understanding of the
canonical connection $\Gamma$; we define the space of surfaces $\T^2_1$
related to the computation of $D_\mu \ket{\wh\O_\nu} -D_\nu \ket{\wh\O_\mu}$,
and study its properties. We also simplify the computation of curvature.
In sect.5 we define our moduli spaces of surfaces with one or more special
punctures and introduce the relevant antibracket, $\K$ operator and $\I$
operator. We also discuss the homomorphism
from moduli spaces to functions on the state space of the conformal theory.
In sect.6 we use our
previous developments to show that the consistency conditions can be
satisfied through the existence of suitable moduli spaces $\B^2$ of 
surfaces with two special punctures. Sect.7 is devoted to the 
recursive construction of the $\B^2$ spaces.

\chapter{Review and Notation}

In this section we will review some of the basic facts required 
in the present work. Most of these facts were explained in detail
in earlier publications so we will not offer derivations. The reader
should have no problem in deriving the quoted results using the
original papers. We begin with connections, and later turn to 
moduli spaces of Riemann surfaces, in particular we consider
the string vertices $\V$
and the $\B^1$ spaces.

\section{Connections on spaces of conformal theories}

Given a space $M$ of conformal theories, with coordinates $x^\mu$ we construct
a vector bundle over $M$ by considering at every point of $M$ a vector
space spanned by a basis 
$\ket{\Phi_i}$ for the states of the conformal theory. 
We use $\psi^i$ to denote a coordinate
on the vector space. A connection on this vector bundle is defined by
coefficients $\Gamma_{\mu i}^{~j}$, and covariant derivatives of functions
on this bundle are defined as
$$D_\mu (\Gamma) A \equiv  \partial_\mu A - 
{\partial_r A \over \partial \psi^i}
 \Gamma_{\mu j}^i \, \psi^j \,.\eqn\dcovd$$
It then follows that
$$[ D_\mu, D_\nu ] A = - {\partial_r A \over \partial \psi^i}\,
 R_{\mu\nu\,j}^{~i }\psi^j =   - {\partial_r A \over \partial \psi}\,
 R_{\mu\nu}\,\psi \,, \eqn\curvd$$
where in the last step we  use matrix notation 
$R_{\mu\nu} = \ket{\Phi_j} \, R_{\mu\nu i}^{~j}\, \bra{\Phi^j}$.  We
have   
$$ R_{\mu\nu} = \partial_\mu \Gamma_\nu- \partial_\nu \Gamma_\mu +
\Gamma_\mu\Gamma_\nu - \Gamma_\nu\Gamma_\mu \,, \eqn\curvdd$$
where we also use matrix notation for the connection
$\Gamma_\mu \equiv \ket{\Phi_j} \, \Gamma_{\mu i}^{~j}\, \bra{\Phi^j}$.
The vector bundles we are interested in have a symplectic form 
$$\bra{\omega_{12}} = -\,{}_1\bra{\Phi^i} \omega_{ij}(x) {}_2\bra{\Phi^j}\,
= \bra{R'_{12}} c_0^{-(2)} \,, \eqn\sympf$$
which satisfies $\bra{\omega_{12}}= - \bra{\omega_{21}}$, and whose
inverse $\ket{S_{12}}$ satisfies $\ket{S_{12}} =\ket{S_{21}}$. Given the
symplectic form we define an antibracket
$$\{ A , B \} = {\partial_r A \over \partial \psi^i}\,\, \omega^{ij}\,
{\partial_l B \over \partial \psi^j} \,, \eqn\antibd$$
with standard exchange property $ \{ A , B\} = - (-)^{(A+1)(B+1)} \{ B, A\}$
and satisfying the Jacobi identity
$$(-)^{(A+1) (C+1) } \{ \{ A, B \} , C \}  \, + \, 
\hbox{Cyclic}(A\to B\to C) = 0\,.\eqn\jacobi$$
It is sometimes convenient to use explicit string field notation $\ket{\Psi}
\equiv \ket{\Phi_i} \psi^i$. In this notation  \dcovd\ reads
$$D_\mu A = \partial_\mu A - {\partial A\over \partial \ket{\Psi}} \,
 \Gamma_\mu \ket{\Psi}\,. \eqn\dcovdn$$
The covariant derivative of functions of the type 
$\bra{A}\Psi\rangle \cdots \ket{\Psi}$ defined by tensor sections $\bra{A}$ is
given by
$$D_\mu \bra{A}\Psi\rangle\cdots \ket{\Psi} = 
(\, D_\mu \bra{A}\, ) \ket{\Psi}\cdots \ket{\Psi}\,,   \eqn\dtensec$$
where covariant derivatives of sections are simply given by 
$$D_\mu (\Gamma) \bra{A} = \partial_\mu \bra{A} - 
\sum_n\bra{A} \Gamma_\mu^{(n)} \,,  \eqn\convs$$
with the label $n$ referring to a state space in the tensor section.
The antibracket can also be written as
$$ \{ A, B\} = (-)^{B+1} {\partial A\over \partial \ket{\Psi}}
{\partial B\over \partial \ket{\Psi}} \ket{S}\,. \eqn\antibnn$$
The  connections we will be dealing with are {\it symplectic}; namely
they satisfy 
$ D_\mu (\Gamma) \bra{\omega_{12}} = 0$. It then follows that covariant
derivatives act as derivations of the antibracket
$$D_\mu \{ A, B\}=\{ D_\mu A\, ,\,B\}+\{ A\, ,\,D_\mu B\}\,.\eqn\cdbra$$
The curvature  of a  symplectic connection
is hamiltonian, namely,  the right hand side of 
\curvd\ can be reproduced by canonical action. Indeed, using
$ \bra{\omega_{12} } ( R_{\mu\nu}^{(1)} +  R_{\mu\nu}^{(2)} ) = 0$, which
follows by commutator action on the symplectic form, one can readily show
that 
$$[D_\mu , D_\nu]\, A = -\,\{ A \, , R_{\mu\nu} \}\,, \eqn\curvf$$
where the curvature hamiltonian $R_{\mu\nu}$, denoted with a slight
abuse of notation with the same symbol as the curvature matrix, is given by
$$R_{\mu\nu} = -\half \bra{\omega_{12}} R^{(2)}_{\mu\nu} \ket{\Psi} 
\ket{\Psi}\,.  \eqn\curvham$$

Having reviewed general facts about connections let us recall
some properties of the particular connection we shall be using here.
This is the connection $\wh\Gamma$ of Ref.[\rangasonodazw], anticipated
in Refs[\campbell,\ranganathan,\kugozwiebach].  This connection,
denoted hereafter as $\Gamma$, is best described 
(in the spirit of [\sonoda]) by indicating how to
take covariant derivatives.
Let $\Sigma$ denote a punctured Riemann surface with punctures
$P_1, \cdots P_n$, with non overlapping coordinate disks $D_1, \cdots D_n$,
and let $\widehat\Sigma = \Sigma - \cup_i D_i$ denote the surface minus
its unit disks. The covariant derivative of the surface state $\bra{\Sigma}$
is given by  
$$D_\mu\bra{\Sigma~}= -{1\over \pi}\hskip-10pt\int_{~~~p\,\in \,\widehat\Sigma} 
\hskip-8pt d\mu(p) \bra{\Sigma ; p\,} \O_\mu (p)\rangle\,. \eqn\conder$$
Here the measure of integration is unambiguous since the operator 
$\O_\mu$ is a dimension $(1,1)$ primary operator.
The covariant derivatives of the Virasoro operators allow one to show that
([\rangasonodazw], Eqn.(5.10)) 
$$\Gamma^k_{\mu\nu} ={1\over \pi}\,
 {  H_{\mu\nu}^k \, \delta_{0,s_k} \over 2-\gamma_k} 
 \,\,, \,\,\hbox{for}\,\,
\gamma_k \not= 2 \,,\eqn\ghty$$
where $s_k$ and $\gamma_k$ denote the spin and total conformal dimension
of the state $\ket{\Phi_k}$. Ref.[\rangasonodazw] also discussed another
connection $c_{\mu i}^{~j}\,$,  first introduced in the work of 
Sonoda [\sonoda]. For 
$\gamma_k \leq 2$ one can show that $\Gamma_{\mu\nu}^{~k} = c_{\mu\nu}^{~k}$
(Ref.[\rangasonodazw], Eqn.(4.10)). This, together with the torsion
analysis of Ref.[\rangasonodazw] (see Eqns.(6.25--29)) which 
explained that $c^{k}_{\mu\nu}- c^{k}_{\nu\mu} = 0$ for $\gamma_k\leq 2$,
implies that
$$\Gamma^{k}_{\mu\nu}-
\Gamma^{k}_{\nu\mu}  = 0 \,,\quad  \hbox{for}\,\,\, \gamma_k \leq
2\,.\eqn\bef$$
It may be possible to give a somewhat simpler derivation of this
fact in the framework of [\gussichsundell].

\section{Moduli spaces of string vertices and $\B$ spaces}
The string vertices $\V = \sum_{n\geq 3} \V_{n}$ at genus zero satisfy
the recursion relations (see, for example, Ref.[\senzwiebachtwo] Eqn.(2.22))
$$\partial \V +  \, \half\{ \V ,\V\} =0\,. \eqn\recrel$$
These recursion relations can be expressed as the fact that the operator 
$\delta_\V$ squares to zero:
$$\delta_\V^2 =0\, ,\quad \hbox{with}\quad
\delta_\V \equiv \partial + \{  \V\, \, ,\,\, \} \eqn\anuop$$
Moreover, we readily find that
$$ \delta_\V \V = \half \{ \V \,, \, \V \} \,. \eqn\neweq$$

It is useful to introduce a new piece of notation, an operator $\I$ that
will be discussed in detail in sect.5. This operator makes one of the 
ordinary punctures of a surface special. Since our moduli spaces are
symmetric under the exchange of labels of punctures on the surfaces, any
puncture may be chosen. This operator acts as a derivation of the antibracket,
and one readily verifies that  
$$ \delta_\V \,\I\V = 0\,. \eqn\consd$$

At genus zero, the interpolating $\B^1$ spaces of background independence
(formerly called $\B$) have one special puncture, and exist for 
two or more ordinary punctures. We write $\B^1\equiv\sum_{n\geq 2}\B^1_{n}$,
where $n$ denores the number of ordinary punctures.
Their recursion relations take the form (see, for example
Ref.[\bergmanzwiebach], Eqn.(3.18))
$$\p\B^1\, = \V'_3 + (\K - \I )\V - \{ \V,\B^1\}\,. \eqn\recb$$
The two lowest dimension cases here are
$$\eqalign{ \partial \B^1_2 &= \V'_3 - \I\V_3 \,, \cr
\p \B^1_3 &= \K\V_3 - \I \V_4 - \{ \V_3 \,,\, \B^1_2\}\, . \cr} \eqn\nytur$$
The recursion relation \recb\ can be written more compactly using the
operator $\delta_\V$ introduced in \anuop. We get 
$$\delta_\V\B^1\, = \V'_3 + (\K - \I )\V
\,\,. \eqn\recbb$$

\def\H{\widehat{\cal H}}
\def\p{\partial}
\def\ws{{\widehat\Sigma}}
\def\wh{\widehat}

\chapter{\bf Commutator of Background Deformations}

In this section we reconsider the equation which defining the 
hamiltonian that implements an infinitesimal background deformation.
For simplicity we only consider the
classical string field theory. The equation in question was derived
in [\senzwiebach] and reads
$$ D_\mu (\,  \Gamma \, )\, S = \{ S\, , \, B_\mu \} \,.\eqn\one$$
Local background independence at the classical level was established
by finding the explicit hamiltonian $B_\mu$ (formerly called $U_\mu$)
satisfying this equation. It was found that $B_\mu$ is given by 
$$B_\mu = B^{(2)}_\mu - f_\mu (\B^1) \,,\eqn\hamback$$
where $B^{(2)}_\mu$ is a quadratic hamiltonian implementing an
$\wh{\cal O}_\mu$ insertion. We will take a
second covariant derivative of \one\ and will show the existence of 
an $S$-closed function $H_{\mu\nu}$. We explain 
how the expected uniqueness of the
string master action requires that $H_{\mu\nu}$ be $S$-trivial:
$H_{\mu\nu} = \{ S , B_{\mu\nu}\}$, with $B_{\mu\nu}$ a hamiltonian
to be determined.   

\section{The commutator conditions}

Taking a second covariant  derivative of Eqn.\one, and making the
connection implicit we find
$$ D_\mu  D_\nu \, S = \bigl\{ \,  \{ S\, , \, B_\mu \}\, , \, B_\nu \bigr\}
+  \{ S\, , \,D_\mu\,  B_\nu \} \,,  \eqn\two$$
where use was made of the fact that the connection is symplectic and therefore
covariant derivatives act as derivations of the antibracket   (see
Eqn.\cdbra). We now form the commutator and find
$$[ D_\mu \,, \,  D_\nu \,]\, S = \Bigl\{ \,\,  S\, , \, D_\mu B_\nu
 -  D_\nu B_\mu +  \{ B_\mu  ,   B_\nu \, \bigr\} \,\Bigr\}  
  \,.  \eqn\three$$
On the other hand the commutator is given by curvature 
$[ D_\mu \,, \,  D_\nu \,]\, S = - \bigl\{ \,\,  S\, , \, R_{\mu\nu} 
\, \bigr\}$ (see \curvf), and therefore we find the consistency condition 
$$\bigl\{ \,\,  S\, , \, H_{\mu\nu}\,\bigr\}=0 \,, \eqn\six$$
where
the object $H_{\mu\nu}$ is given by
$$ H_{\mu\nu} \equiv  D_\mu B_\nu
 -  D_\nu B_\mu +  \{ B_\mu  ,   B_\nu \} +\, R_{\mu\nu}\,.\eqn\four$$
It should be emphasized that this consistency condition is clearly
satisfied,
since it follows from Eqn.\one, an equation that has been solved explicitly.
Nevertheless the hoped for uniqueness of the string master action $S$ 
satisfying the master equation $\{ S , S \}=0$, implies that a perturbed
master action $S+ \lambda^{\mu\nu} H_{\mu\nu}$, which would
satisfy the master equation to first order in $\lambda$, should 
just be a field redefined
version of the original action. For this to be the case we must demand
that there is a hamiltonian $B_{\mu\nu}$ such that 
$$ H_{\mu\nu} =  
\bigl\{ \,\,  S\, , \, B_{\mu\nu} \, \bigr\} \,.  \eqn\seven$$
If \seven\ holds then \six\ will follow from the Jacobi identity.
More explicitly Eqn.\seven\ reads
$$  D_\mu B_\nu -D_\nu B_\mu +\{ B_\mu  , B_\nu \}\, +\,  \, R_{\mu\nu}=  
\bigl\{ \,\,  S\, , \, B_{\mu\nu} \, \bigr\} \,. \eqn\eight$$
This is the equation we will analyze. The existence of $B_{\mu\nu}$
amounts to a (partial) cohomology theorem for the string action. It asserts
that the $S$-closed hamiltonian $H_{\mu\nu}$ is actually $S$-exact, it 
equals $\{ S, B_{\mu\nu}\}$.
Moreover, it will be
seen that the hamiltonian
$B_{\mu\nu}$ is defined by moduli spaces of surfaces with two special 
punctures.

We can now understand in retrospect that even the existence of the original
hamiltonian $B_\mu$ of \one\ amounts to a cohomology theorem. Since we use
symplectic connections it follows from $\{ S, S\} =0$ that 
$\{ S, D_\mu S \} =0$. Equation \one\ simply says that the $S$-closed
function $D_\mu S$ is actually $S$-exact.

\section{Preliminary Analysis}

Let us explore the uniqueness of the object $B_{\mu\nu}$ we are after.
In equation \one\ the connection $\Gamma_\mu$ is a reference 
symplectic connection, and can be shifted as long as we preserve the
symplectic nature of the connection. Let  
$\Delta \Gamma_\mu^{(i)}$   denote the matrix encoding the change in
the connection, and $\Delta \Gamma_\mu$ the associated hamiltonian
$$\Delta \Gamma_\mu =  -\half \bra{\omega_{12}} \Delta \Gamma_\mu^{(2)}
\ket{\Psi}\ket{\Psi} \eqn\nine$$
We now claim that
$$\eqalign{\Gamma_\mu \,&\to \, \Gamma_\mu + \Delta \Gamma_\mu \,,\cr 
B_\mu \,&\to \, B_\mu - \Delta \Gamma_\mu \,, \cr } \eqn\ten$$
leaves invariant the background independence condition \one. 
This is a simple consequence of
$$D_\mu (\, \Gamma + \Delta\Gamma \, ) A = D_\mu (\, \Gamma \, ) A 
+ \{ \Delta \Gamma\,,\, A \, \} \,,\eqn\eleven$$
which holds for any arbitrary function $A$.

We can now see that the shift \ten\ does not alter the determination of
$B_{\mu\nu}$ in Eqn.\eight. 
A short computation gives
$$\eqalign{
 D_\mu B_\nu -D_\nu B_\mu +\{ B_\mu  , B_\nu \}\, &\to \,\,
 D_\mu B_\nu -D_\nu B_\mu +\{ B_\mu  , B_\nu \}\,\cr
&\quad  -  D_\mu\,\Delta\Gamma_\nu +
D_\nu\,\Delta\Gamma_\mu - \{ \Delta\Gamma_\mu\, 
 , \Delta\Gamma_\nu \} \,,  \cr}  \eqn\varcur$$
and for the curvature we find
$$R_{\mu\nu} \,\to \, R_{\mu\nu}  + D_\mu\,\Delta\Gamma_\nu -
D_\nu\,\Delta\Gamma_\mu + \{ \Delta\Gamma_\mu\, 
 , \Delta\Gamma_\nu \}\,. \eqn\thirteen$$
It follows immediately that $H_{\mu\nu}$ is left
invariant under a shift of the symplectic connection. This proves
that the choice of symplectic
connection is irrelevant to the computation of
$B_{\mu\nu}$. We can therefore use the canonical unit-disk connection $\Gamma$
in the evaluation of $H_{\mu\nu}$ and in the subsequent computation of
$B_{\mu\nu}$.
\bigskip
Let us now consider a different type of transformation.
Equation \one\ is clearly invariant under the shift
$$B_\mu \, \to \, B_\mu + \{ S \, , \, \lambda_\mu \} \,.\eqn\fourteen$$
Under this shift $H_{\mu\nu}$ transforms as
$$H_{\mu\nu} \,\to \, H_{\mu\nu} + \{ \, S\, , \D_\mu \lambda_\nu - 
\D_\nu \lambda_\mu \, \}\,, \eqn\seventeen$$
where we have defined a ``gauge covariant" derivative
$${\cal D}_\mu \equiv D_\mu + \{  B_\mu \, , \, \, \}  \,. \eqn\fifteen$$
It follows that  under the shift \fourteen\ our solution for $B_{\mu\nu}$
will shift by 
$$B_{\mu\nu} \,\to \, B_{\mu\nu} + \D_\mu \lambda_\nu - 
\D_\nu \lambda_\mu \,\,. \eqn\eighteen$$
This is a true non-uniqueness of $B_{\mu\nu}$.
It might be worth remarking that the gauge-covariant derivative 
introduced above affords some nice notation. 
Eqn.\one\ reads
${\cal D}_\mu S = 0$, and
$ [ \D_\mu \, , \D_\nu ] = -\, \{ \, H_{\mu\nu} \, , \,\,\,\, \} $.

\bigskip
\chapter{Properties of Connections and Covariant Derivatives}

In this section we find the geometrical meaning
of the antisymmetric combination $D_\mu\ket{\wh\O_\nu}- D_\nu\ket{\wh\O_\mu}$.
This object will be seen to have a representation in terms of
a moduli space $\T^2_1$ of  spheres with three punctures, two of which are
special. This moduli space is defined to have real dimension one, but
in some sense its natural dimension is two. Since it is a moduli space
of three punctured spheres, and the position of three punctures on a sphere
define no moduli, the moduli parameters of this moduli space refer to the
coordinates at the punctures. Nevertheless it is simpler to think of the
punctures as moving, carrying along some canonical coordinates. The
derivation of these results will take place in the first three subsections.
The last subsection gives a simplified proof of the fact that the connection
$\Gamma$ has zero curvature [\rangasonodazw]. This is possible due to the
geometrical understanding of the antisymmetric combination mentioned above.

\section{Antisymmetric Derivatives}

The purpose of the present subsection is to establish the following result:
$$D_\mu \ket{\O_\nu} - D_\nu \ket{\O_\mu} = - \,{1\over \pi}
\int_{|z|<1} \hbox{dxdy}
\Bigl( \, \O_\mu (z,\bar z)  \, \ket{\O_\nu }
- \O_\nu (z, \bar z)  \, \ket{\O_\mu } \Bigr) \,.\eqn\torsc$$
This equation indicates
that the antisymmetric combination of covariant derivatives
using the connection $\Gamma$ is simply related to integrated correlators,
and therefore will allow us to understand the antisymmetric combination as 
arising from a moduli space of surfaces. Since the moving puncture
in the right hand side spans a space of real dimension two, the relevant
moduli space can be thought to be of dimension two. Nevertheless,
since the rotation $z\to ze^{i\theta}$ is naturally incorporated in
our formalism, it will turn out convenient to think of the moduli space
as having one real dimension only. Since the connection $\Gamma$
acts trivially on the ghosts the above equation will imply that
$$D_\mu \ket{\wh\O_\nu} - D_\nu \ket{\wh\O_\mu} =
-\, {1\over \pi}\int_{|z|<1} \hbox{dxdy}
\Bigl( \, \O_\mu (z,\bar z)  \, \ket{\wh\O_\nu }
- \O_\nu (z, \bar z)  \, \ket{\wh\O_\mu } \Bigr)\,, \eqn\gtor$$
where $\ket{\wh\O} \equiv \ket{c\bar c \O}$.
\medskip
Let us now give a proof of Eqn.\torsc. By definition of the connection
coefficients we have
$$D_\mu \ket{\O_\nu} - D_\nu \ket{\O_\mu}= \sum_{\gamma_k} 
(\, \Gamma^{k}_{\mu\nu}-
\Gamma^{k}_{\nu\mu} ) \ket{\Phi_k}\,, \eqn\enough$$
and using \ghty, and \bef, we find 

$$D_\mu \ket{\O_\nu} - D_\nu \ket{\O_\mu} = 
-\, {1\over \pi} \sum_{\gamma_k>2} { H_{\mu\nu}^k-H_{\nu\mu}^k \, \over \gamma_k - 2}
\,\delta_{0,s_k} \ket{\Phi_k} \,. \eqn\frst$$
On the other hand from the definition of operator product,
and when integrating over
a domain with rotational symmetry:
$$\O_\mu (z,\bar z)\,\O_\nu (0) \ket{0} ={1\over 2\pi} \sum_{\gamma_k}
{ H_{\mu\nu}^k \,\delta_{s_k,0}\over r^{4-\gamma_k}}\ket{\Phi_k}\,. \eqn\opp$$
We use this equation to form the antisymmetric operator product
$$\Bigl( \, \O_\mu (z,\bar z)  \, \O_\nu (0) 
- \O_\nu (z, \bar z)  \, \O_\mu (0) \Bigr) \ket{0} = {1\over 2\pi} 
\sum_{\gamma_k >2}
{( H_{\mu\nu}^k- H_{\nu\mu}^k) \,\over r^{4-\gamma_k}  } \delta_{s_k,0}\,
 \ket{\Phi_k}  \eqn\antsym$$
In the right hand side we have restricted the sum to 
$\gamma_\k >2$ following the discussion of Ref.[\rangasonodazw] (see below
Eqn.(6.23)) where it is shown that the left hand side of \antsym\ is
integrable over the disk and therefore conformal fields with lower dimension
cannot appear. Finally, integrating over the unit disk $|z| < 1$ we find
$$\int_{|z|<1} \hbox{dxdy}
\Bigl( \, \O_\mu (z,\bar z)  \, \O_\nu (0) 
- \O_\nu (z, \bar z)  \, \O_\mu (0) \Bigr) \ket{0} = \sum_{\gamma_k >2}
{( H_{\mu\nu}^k - H_{\nu\mu}^k )\, \over \gamma_k-2   }\,\delta_{s_k,0}\, 
\ket{\Phi_k}\,. \eqn\inta$$
Comparing this equation with \frst\ we see that indeed \torsc\ has been
established. In order to associate a moduli space to the right hand side of
\torsc\ we now set up some notation to deal with three punctured spheres
and local operators.  

\section{Standard Spheres and Local Operators}

We consider three punctured spheres where the underlying unpunctured
sphere is described by uniformizers $z$ and $w$ satisfying
$zw=1$. Two of the punctures of the sphere will be considered 
ordinary; the first, labelled `$1$', will be located at $z=0$, 
with local coordinate
$z_1 = z$, and the second, labelled `$2$' will be located at $w=0$,
 with local coordinate $w_1 = w$. 
The third puncture will be thought as special, and will be labelled `$\bar 1$'.
Let $S(1)$ denote the sphere 
$$S(1) : \,\,  z_1 = z, \quad w_1 = w, \quad w_{\bar 1} = w-1\,. \eqn\sphr$$
This sphere is equivalent to the sphere  $\V'_3$ relevant appearing in \recb,
up to the definition of the local coordinate at the special puncture, which
does not matter in many cases.
Let now $S(t)$ denote a sphere with the special puncture at $w=t$:
$$S(t) : \,\,  z_1 = z, \quad w_1 = w, \quad w_{\bar 1} = w-t \,. \eqn\sphrp$$
Given the conformal equivalence of all three punctured spheres, $S(t)$
differs from $S(1)$ only by a change of local coordinates at the 
punctures.  In fact, one can readily verify that $S'(1)$ defined by the
following change in the local coordinates of $S(1)$ 
$$S'(1) : \,\,  \tilde z_1 = z_1/t,  \quad \tilde w_1 = t \,w_1, \quad 
\tilde w_{\bar 1} = t\, w_{\bar 1}\,, \eqn\nhg$$
is conformally equivalent to $S(t)$. All spheres $S(t)$ just differ
by scalings of their local coordinates. Nevertheless the parameter $t$ is
most easily thought of as specifiying a position.
A scaling of the local coordinate on a surface state is realized as
$\bra{\Sigma ; {w_1\over t}} = \bra{\Sigma ; w_1} \, t^{L_0^{(1)}}
 {\bar t}^{\bar L_0^{(1)}}$, and therefore we can relate the sphere
surface states as follows
$$\bra{S(t)} = \bra{S(1)} [ t^{L_0^{(1)} } \bar t^{\bar L_0^{(1)} }] 
[ t^{-L_0^{(2)} } \bar t^{-\bar L_0^{(2)} }]
[ t^{L_0^{(\bar 1)} } \bar t^{\bar L_0^{(\bar 1)} }] \,,\eqn\mpsph$$
where the Fock space label $(1)$ is attached to the puncture at $z=0$,
the label $(2)$ is attached to the puncture at $w=0$, and the label
$(\bar 1)$ is attached to the special puncture.

The above spheres actually define what we usually mean by local operators.
One denotes by
$\Phi(t)$ the operator constructed by inserting the state $\ket{\Phi}$
on  the special puncture of the sphere $S(t)$, and using a reflector
to turn one of the state spaces on the surface state bra into a ket:
$${}_{2'}[ \Phi (t)]_2=\bra{S(t)}\Phi\rangle_{\bar 1}\ket{R_{12'}}\,.\eqn\prt$$
Using this definition and \mpsph\ we derive the operator relations
$$\eqalign{
\Phi(z) &= {1\over z^h \bar z^{\bar h}} \,\,\, z^{L_0} \bar z^{\bar L_0}
\, \Phi(1)\, z^{-L_0} \bar z^{-\bar L_0} \,, \cr
\Phi(tz) &= {1\over t^h \bar t^{\bar h}} \,\,\, t^{L_0} \bar t^{\bar L_0}
\, \Phi(z)\, t^{-L_0} \bar t^{-\bar L_0}\,.\cr} \eqn\opmap$$

\section{Definition and Properties of $\T^2_1$}

In this section we will introduce a moduli space $\T^2_1$ related to
Eqn\torsc. This will be a moduli space of spheres with three punctures,
two of which will be special. The real dimension of this space will
be one. 

We begin by introducing for any $0<u<1$ a space $\tau_1^2 (u)$
of three punctured spheres. The space
is of one real dimension and is parametrized by $t$ as follows
$$\tau_1^2(u) = \Bigl\{ S(t)\,  \bigl|\,  t \in [u,1]  \Bigr\}\,,
 \eqn\onedim$$
where $S(t)$ was defined in \sphrp. 
The surfaces in the moduli space $\tau^2_1(u)$ are defined to have 
two special punctures
(thus the superscript value) and one ordinary puncture (thus the
subscript value). The special punctures are
the moving puncture which was  labelled as $\bar 1$, 
and the puncture at $w=0$, which used to be the puncture $2$, but
now will be  labelled $\bar 2$. The moving
puncture moves from the point $w=u$ up to the point $w=1$.
When the moving puncture starts its journey it is closest to the
second special puncture.

\Figure{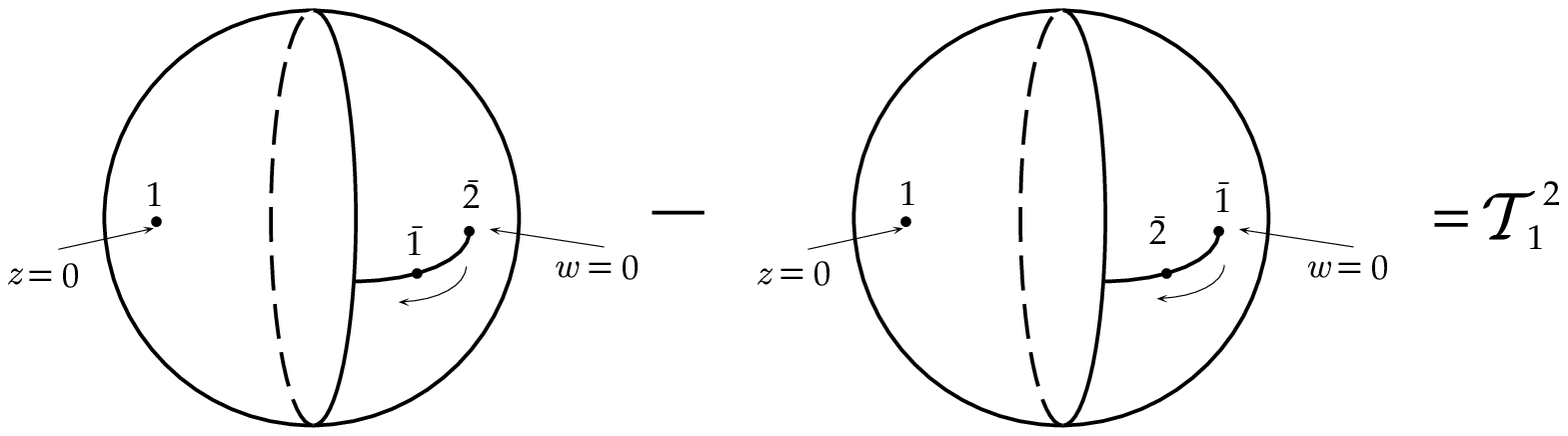}{Figure 1.~The family of three punctured spheres that defines the
space $\T^2_1$. This space has two special punctures appearing 
antisymmetrically. One of the special punctures travels from the point 
$w=0$, where it coincides with the other special puncture, up to the
point $w=1$.}

It is useful to consider the  limit where
we take $u\to 0$ and the moving puncture starts closer and closer
to the second special puncture.  The limit
$\lim_{u\to 0} \tau^2_1 (u)$, however, cannot be expected to be well
defined
since the
operator product expansion of the two marginal 
states to be inserted at the special
punctures will not give an integrable expression 
(in the parameter $t\in [0,1]$). 
We must antisymmetrize on the special punctures
and define a moduli space $\T^2_1(u)$ with two special punctures
as:  
$$\T^2_1 (u) = \tau^2_1(u)  - P \tau^2_1(u) \,,\eqn\intdefd$$
where $P$ simply
exchanges the labels $\bar 1$ and $\bar 2$ of the special punctures.
Moreover, we can now define the limit
$$\T_1^2 \equiv  \lim_{u\to 0} \T^2_1 (u)\,,  
 \eqn\odim$$   
which makes sense because antisymmetrized operator products of 
marginal operators should be integrable (see Figure 1). It follows
from the definition \intdefd\ that
$$ \partial \T^2_1(u)  =  \partial\, ( \, \tau^2_1(u)  - P \tau^2_1(u)  \,)
= ( \V'_3 - P \V'_3 ) - ( S (u) - P S(u) ) \,,\eqn\bdtau$$
where we recall that $\V'_3 = S(1)$ except for the fact that $\V'_3$
is now considered to have two special punctures.
Consider now taking now the limit when $u\to 0$.  The antisymmetric combination 
$( S (u) - P S(u) )$ must be taken to vanish since the punctures that
are being antisymmetrized are getting close to each other and thus 
becoming progressively symmetric under their exchange.  We then find
$$ \partial \T^2_1 =  \I \V'_3 \,,\eqn\boundtau$$ 
where anticipating the notation to be discussed in the next section, the
operator $\I$ turns an ordinary puncture special, and antisymmetrizes
the resulting space in the labels of the special punctures.

We now turn to the derivation of a relation between two dimensional insertions
and insertions via $\T^2_1$.  We use the notation
$$f_{\O_i\O_j}( \tau_1^2)  \equiv \int_{\tau_1^2} \bra{\Omega_{1 \bar 1 \bar 2}} 
\O_i\rangle_{\bar 1}
\ket{\O_j}_{\bar 2} \ket{\Psi}_1\,, \eqn\aa$$
where the canonical CFT-valued moduli space form $\bra{\Omega}$
is the usual form used in string field theory (see, for example
Ref.[\senzwiebachtwo], Eqn.(3.10)).
The states appearing as subscripts in $f$ are inserted 
on the special punctures, 
 $\O_i$ on the punctured labelled ${\bar 1}$,  and  $\O_j$ on the 
puncture labelled $\bar 2$.
We now claim that 

\noindent
\underbar{Theorem}
$$ \bra{\omega_{12}}\,\, {1\over \pi}\hskip-6pt \int_{u<|z|<1} 
\hskip-5pt\hbox{dxdy} 
\, [\O_\mu (z,\bar z) ]_2 \, |\wh
\O_\nu\rangle_2 \ket{\Psi}_1 = f_{\mu\nu} (\tau^2_1(u)) \eqn\bb$$
where the state space $(2)$ is associated to the unit disk $|z|<1$, and 
$\wh\O_\nu$ is inserted at $z=0$. In the right hand side, the 
symbol $f_{\mu\nu}$ stands for $f_{\wh\O_\mu \wh\O_\nu}$.

\noindent
\underbar{Proof}~
Begin with the left hand side. Using \opmap\ we write 
$$\O_\mu (z, \bar z)
= {1\over z\bar z}\,\, z^{L_0} \bar z^{\overline L_0}
\, \O_\mu (1) \, z^{-L_0} \bar z^{-\overline L_0}\,, \eqn\cc$$
and using $z=re^{i\theta}$, the left-hand side becomes, 
$$\hbox{LHS} = 
{1\over \pi} \, (2\pi)  \bra{\omega_{12}}\Bigl(\, \int_u^1 \hbox{rdr} 
\,\hbox{r}^{L_0+\overline
L_0-2} \int_0^{2\pi} 
{d\theta\over 2\pi}
 \,e^{i\theta (L_0-\overline L_0)} \O_\mu (1)
 z^{-L_0} \bar z^{-\overline L_0} 
 \, |\wh\O_\nu\rangle \Bigr)_2 \ket{\Psi}_1\,. \eqn\dd$$ 
Since $L_0\ket{\O_\nu} =  \overline L_0\ket{\O_\nu} =0$, and
$\bra{\omega_{12}} (L_0-\overline L_0)_2 = 0$, we find
$$\hbox{ LHS} =  2 \bra{\omega_{12}}\Bigl( \int_u^1 \hbox{dr} 
\,\hbox{r}^{L_0+\overline
L_0-1}  \O_\mu (1)\, |\wh\O_\nu\rangle \Bigr)_2 \ket{\Psi}_1 \,\,.\eqn\lhsc$$ 

\noindent
Now consider the right hand side (RHS), which by definition
(\aa) is the integral of the canonical string one-form over the
space $\tau_1^2 (u)$
$$\hbox{RHS} = \int_{\tau_1^2 (u) }  \bra{\Omega_{1\bar 1 \bar 2}^{[1]} } 
\wh\O_\mu\rangle_{\bar 1}
|\wh\O_\nu\rangle_{\bar 2} \ket{\Psi}_1\,. \eqn\ee$$
Since $\tau_1^2 (u)$ is a moduli space of three punctured spheres 
the canonical normalization factor of the string form equals one, and
the string form is simply the surface state with the appropriate
antighost insertions.  We have (using $z=t+w = h_t(w)$ in the setup
of Ref.[\bergmanzwiebach], Eqns.(2.30--33)) 
$$\eqalign{
\bra{\Omega_{1\bar 1\bar 2}^{[1]} } \wh\O_\mu\rangle_{\bar 1} 
&= \,\,\bra{S(t)_{1\bar 1\bar 2}}\, dt\, {\bf b} \bigl(
{\partial\over \partial t}\,\bigr) |\wh\O_\mu\rangle_{\bar 1} \,\,,\cr
&= - dt\,\bra{S(t)_{1\bar 1\bar 2}}\, \Bigl( (b_{-1} + \overline b_{-1}) 
c_1\overline c_1 | \wh\O_\mu\rangle \Bigr)_{\bar 1}\,\,,\cr
&=   dt\,\bra{S(t)_{1\bar 1\bar 2}}\, \Bigl( (c_{1} - \overline c_{1}) 
 |\O_\mu\rangle \Bigr)_{\bar 1}\,\,,\cr
&= dt\,\bra{R_{1 \bar 2}}\, \Bigl( c\O_\mu (t) - \overline c\O_\mu (t) 
\Bigr)_{\bar 2} \,\,.\cr}  \eqn\ff$$
Now we use ($t = z = \bar z)$ 
$$\Biggl\{ {c\O_\mu (t)\atop \bar c\O_\mu (t)} \Biggr\} 
 = t^{L_0 + \overline L_0 -1}\Biggl\{ { c\O_\mu (1) \atop  \bar c\O_\mu (1) }
\Biggr\} \,t^{-L_0 - \overline L_0}\,,  \eqn\gg$$
to find
$$\hbox{RHS} 
=  \bra{R_{1\bar 2}}\,\Bigl(\,\,\int_u^1 dt
\,t^{L_0 + \overline L_0 -1} 
( c(1) - \overline c (1))   \O_\mu (1)   \, |\wh\O_\nu\rangle \Bigr)_{\bar 2}
\ket{\Psi}_1\,\,. \eqn\ii$$
Making use of  $\ket{\Psi} = b_0^- c_0^-\ket{\Psi}$, the
reflection property
$\bra{R_{1\bar 2}} b_0^{-(\bar 2)} = \bra{R_{1\bar 2}} b_0^{-(1)}$, 
and the anticommutator
$\{ b_0^- , c(1)- \bar c (1) \} = 2$, we find that the right hand side becomes
$$\eqalign{
\hbox{RHS} &= -2\,  \bra{R'_{1\bar 2}}\,\Bigl(\,\,\int_u^1 dt
\,t^{L_0 + \overline L_0 -1} 
 \O_\mu (1)   \, |\wh\O_\nu\rangle \Bigr)_{\bar 2}
c_0^-\ket{\Psi}_1 \,\,, \cr
&=  2\,  \bra{\omega_{1\bar 2}}\,\Bigl(\,\,\int_u^1 dt
\,t^{L_0 + \overline L_0 -1} 
 \O_\mu (1)   \, |\wh\O_\nu\rangle \Bigr)_{\bar 2}
\ket{\Psi}_1\,\,.\cr} \eqn\rhsc$$
Comparison with \lhsc\ establishes the desired result (eqn.\bb). 

Given the definition
$$U^{(2)}_\O = \bra{\omega_{12}} \O\rangle_1 \ket{\Psi_2}\,,   \eqn\sinsert$$ 
we have shown that
$$ U_{-{1\over \pi}\hskip-2pt \int_{u<|z|<1}
\hskip-2pt\hbox{dxdy} 
\, [\O_\mu (z,\bar z) ]\, |\wh
\O_\nu\rangle}^{(2)} = f_{\mu\nu}
 (\tau_1^2(u)) \,.\eqn\aaa$$
Combining this result with \gtor\ we now write
$$ U^{(2)}_{D_\mu|\wh\O_\nu\rangle - D_\nu|\wh\O_\mu \rangle }
 = f_{\mu\nu} (\T_1^2) \,\,,\eqn\bbb$$
where we made use of \intdefd\ and have taken the limit  $u\to 0$ 
which is allowed
on the antisymmetric combination. This last equation expresses the fact
that an insertion of the state 
$D_\mu|\wh\O_\nu\rangle - D_\nu|\wh\O_\mu \rangle$ can be achieved
via the insertion
of the states $\ket{\wh\O_\mu}$ and $\ket{\wh\O_\nu}$ on the moduli space
$\T^2_1$ of three punctured spheres.

\section{Simplifying a curvature computation}

It is now possible to use the results of the previous subsections to
simplify the proof that the connection $\Gamma$ has zero curvature
[\rangasonodazw]. Recall the definition \conder, which for convenience
we reproduce here
$$D_\mu\bra{\Sigma~}= -{1\over \pi}\hskip-10pt\int_{~~~p\,\in \,\widehat\Sigma} 
\hskip-8pt d\mu(p) \bra{\Sigma ; p\,} \O_\mu (p)\rangle\,, \eqn\wellit$$
In order to be able to take a further covariant derivative we must specify
the coordinate disk to be used for the point $p$ as it moves around in
$\wh\Sigma$. Let $\Delta_p$ denote this disk, which we fix 
in such a way that for any puncture $P_i$, as the point $p$ approaches 
the boundary of the disk
$D_i$ the disk $\Delta_p$ does not contain the puncture $P_i$. We can then
take another derivative to find, using the convenient notation
introduced in [\cornalba]
$$\eqalign{
D_\nu\, D_\mu \bra{\Sigma~}  &= 
-{1\over \pi}\hskip-10pt\int_{~~~p\,\in \,\ws } 
\hskip-6pt d\mu(p) \bra{\Sigma ; p\,} 
\,(\,  D_\nu  \ket{\O_\mu (p)} \, )  \cr
  &\quad +{1\over \pi^2}\hskip-6pt\int_{~~~p\,\in \,\ws } 
\hskip-6pt d\mu(p) \hskip-6pt\int_{{q\,\in \,\ws}
\atop {q\,\not\in \Delta_p} } 
\hskip-6pt d\mu(q) \bra{\Sigma ; p,q\,} \O_\mu (p)\rangle \ket{\O_\nu (q)}\cr
  &\quad  -{1\over \pi^2}\sum_i\hskip-6pt\int_{~~~p\,\in \,\ws } 
\hskip-8pt d\mu(p) \hskip-9pt\int_{~~~q\,\in \,D_i\cap \Delta_p } 
\hskip-10pt d\mu(q) \bra{\Sigma ; p,q\,} \O_\mu (p)\rangle 
\ket{\O_\nu (q)}\,\,.\cr} \eqn\tcov$$
The first term in the right hand side is simply computing the covariant
derivative of the state $\ket{\O_\mu}$, while the second and  third 
terms deal with the covariant derivative of $\bra{\Sigma ; p}$. 
When the disk $\Delta_p$ 
lies completely within $\wh\Sigma$ the third term vanishes and the 
second term computes the correct derivative. When the disk $\Delta_p$ 
intersects a disk $D_i$,  the intersection $D_i\cap \Delta_p$ must
be counted negatively, and the third term does this. This is readily
understood: the definition of $\Gamma$ requires integrating over the surface
minus all the coordinate disks
$\Sigma - \cup D_i - \Delta_p = \wh\Sigma-
\Delta_p$. This region, however, can be written as  
$[\wh\Sigma - \wh\Sigma\cap\Delta_p] - 
[\Delta_p -\wh\Sigma\cap\Delta_p]$. The first term in this difference is
taken care by the second term in the above right hand side, while the
second term represents the integral over $\Delta_p\cap D_i$ and counts
negatively, as the third term in the above right hand side does. This last
term actually gives rise to a divergence when the point $p$ approaches the
boundary of the disks $D_i$. This divergence will cancel out in the 
commutator which we now compute
$$\eqalign{
[D_\nu\,, D_\mu \,] \bra{\Sigma~}  &= 
-{1\over \pi}\hskip-6pt\int_{~~~p\,\in \,\ws} 
\hskip-6pt d\mu(p) \bra{\Sigma ; p\,} 
\,(\,  D_\nu  \ket{\O_\mu (p)} \,- D_\mu  \ket{\O_\nu (p)}  )  \cr
  &\quad +{1\over \pi^2}\hskip-6pt\int_{~~~p\,\in \,\ws } 
\hskip-6pt d\mu(p) \hskip-6pt\int_{{q\,\in \,\ws}
\atop {q\,\not\in \Delta_p} } 
\hskip-6pt d\mu(q) \bra{\Sigma ; p,q\,} \Bigl(
\ket{\O_\mu (p)} \ket{\O_\nu (q)} 
-\ket{\O_\nu (p)} \ket{\O_\mu (q)}\Bigr)\cr
&\quad  -{1\over \pi^2}\sum_i\hskip-6pt\int_{~~~p\,\in \,\ws } 
\hskip-8pt d\mu(p) \hskip-9pt\int_{~q\,\in \,D_i\cap \Delta_p } 
\hskip-10pt d\mu(q) \bra{\Sigma ; p,q\,} 
\Bigl( \ket{\O_\mu (p)} \ket{\O_\nu (q)} 
-\ket{\O_\nu (p)} \ket{\O_\mu (q)} \Bigr)\,.\cr} \eqn\ccov$$

\Figure{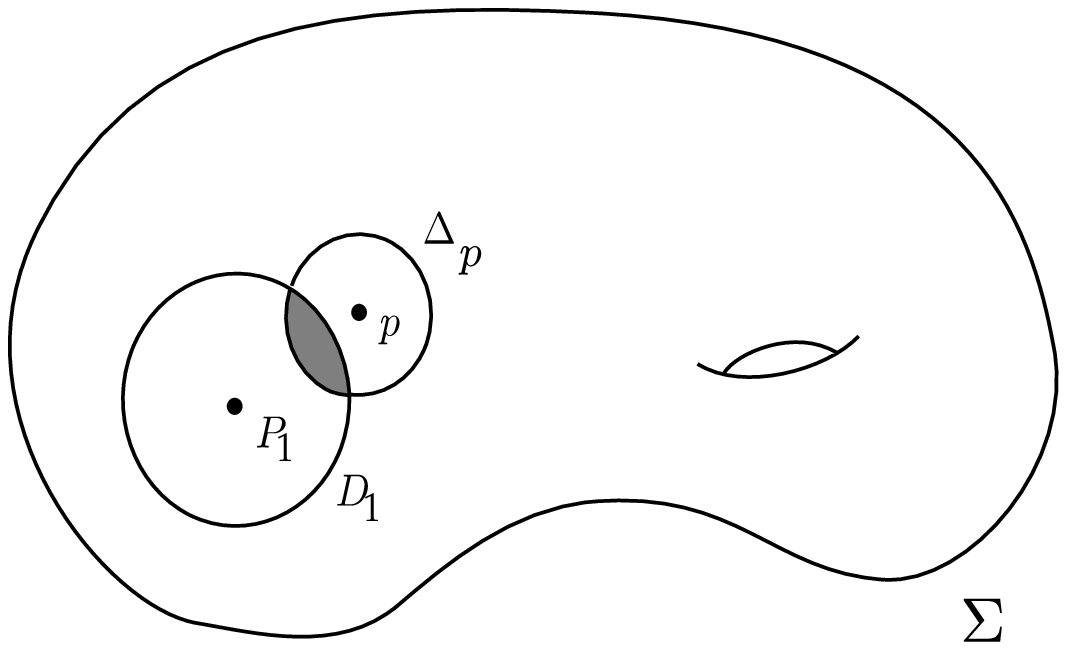 scaled 750}{Figure 2.~This 
figure refers to the computation of curvature for the
connection $\Gamma$. We see the first moving puncture at $p$
and its disk $\Delta_p$ which happens to intersect the unit disk $D_1$ of
the ordinary puncture $P_1$. In this case the shaded region $\Delta_p\cap D_1$
is a region of integration for the second puncture and
contributes with a negative sign.}

Using \torsc\ the first line in the above equation can now be written as
$$ \quad +{1\over \pi^2}\hskip-8pt\int_{~~~p\,\in \,\ws } 
\hskip-8pt d\mu(p) \hskip-8pt\int_{~~~q\,\in \Delta_p}  
\hskip-8pt d\mu(q) \bra{\Sigma ; p,q\,}
\Bigl(  \ket{\O_\mu (p)} \ket{\O_\nu (q)}-
 \ket{\O_\nu (p)} \ket{\O_\mu (q)} \Bigr) \,.\eqn\uhyu$$
Back in \ccov\ we see that the six terms combine naturally into
two groups, each with three terms, one having the states 
$\ket{\O_\mu (p)} \ket{\O_\nu (q)}$ in the integrand, and the
other having the states $\ket{\O_\nu (p)} \ket{\O_\mu (q)}$ in the
integrand.  The first group can be written as
$$ {1\over \pi^2}\,\hskip-14pt\int_{~~~p\,\in \,\ws }
\hskip-12pt d\mu(p) \,\Bigl[ 
  \hskip-8pt\int_{~~~q\,\in \Delta_p}\hskip-12pt  d\mu(q)    
+ \hskip-6pt \int_{{q\,\in \,\ws}
\atop {q\,\not\in \Delta_p} }\hskip-8pt  d\mu(q)   
 -\sum_i \hskip-19pt\int_{~~~~~q\,\in \,D_i\cap \Delta_p }
\hskip-21pt  d\mu(q)  \Bigr]
\bra{\Sigma ; p,q\,}
\O_\mu (p)\rangle \ket{\O_\nu (q)} \,. \eqn\rdtr$$
There are two classes of points $p$. In the first class $\Delta_p$ is
completely within $\ws$, the last integral in the bracket
vanishes and the first two build the integral of $q$ over the complete
$\ws$. In the second class  $\Delta_p$ intersects
a disk $D_i\,$. In this case the first and last integral 
within the bracket build the integral over the part of $\Delta_p$ in $\ws$,
and together with the second integral they again give the integral of $q$
over the complete surface $\ws$.  Thus in both cases the integral over
$q$ extends over $\ws$ and we get 
$$ {1\over \pi^2}\,\hskip-14pt\int_{~~~p\,\in \,\ws }
\hskip-12pt d\mu(p) \,
  \hskip-8pt\int_{~q\,\in\, \ws}\hskip-10pt  d\mu(q)      
\,\bra{\Sigma ; p,q\,}
\O_\mu (p)\rangle \ket{\O_\nu (q)} \,. \eqn\fcom$$
Since the integrations do not distinguish between
$p$ and $q$, the above object
is symmetric under the exchange of $\mu$ and $\nu$. This being the
case it cancels against the other group of three terms which gives
an identical contribution except for the exchange of $\mu$ and $\nu$. 
This shows that the the right hand side of \ccov\ vanishes, and therefore,
that the connection $\Gamma$ has zero curvature. This proof will motivate
our definition of the operator $\K$ in the next section, and the zero 
curvature property
will be related to $\K^2 =0$.

\def\A{{\cal A}}
\def\B{{\cal B}}
\def\K{{\cal K}}
\def\I{{\cal I}}
\def\M{{\cal M}}
\def\O{{\cal O}}
\def\T{{\cal T}}
\def\V{{\cal V}}
\def\H{\widehat{\cal H}}
\def\p{\partial}

\chapter{BV algebra on Riemann Surfaces with Special Punctures}

The purpose of the present section is to extend the definition
of the BV algebra of moduli spaces of punctured Riemann surfaces
[\senzwiebachtwo,\senzwiebachnew]
to the case when the surfaces have special punctures. We will define
a new operator $\I$ and the earlier definition of $\K$ will have to
be extended. Some useful properties of these operators will be 
derived.

For clarity we will denote as $\A$ spaces moduli spaces whose surfaces have
no special punctures.  
Moduli spaces whose surfaces have
a number of special punctures will be denoted as $\B$ spaces. 
An $\A$ space is a particular
case of a $\B$ space with zero special punctures. $\B$ spaces  will
be taken to be symmetric under the exchange of labels of any
pair of ordinary
punctures and antisymmetric under the exchange of labels of any pair of special
punctures. The special punctures are not used for sewing, but the 
antisymmetry property must be preserved under any operation. A $\B$
space will be said to be of type $(n, \bar n)$, and sometimes to be written
as $\B^{\bar n}_n$ when it has $n$ ordinary punctures and $\bar n$ 
special punctures (the bar in $\bar n$ is not meant to imply
any relation to $n$). In a type $(n, \bar n)$ space the ordinary punctures
are labelled from $1$ to $n$, and the special punctures are labelled from
$\bar 1$ to $\bar n$. The ordering of the ordinary punctures is irrelevant,
but the ordering of the special punctures is important for signs to work
out.

A useful notion for a $\B$ space is that of a {\it primitive domain}.
A primitive domain $\overline{\B}$ of a $\B$ space with $\bar n$
special punctures is a subspace of $\B$
such that $\B$ is reproduced by the disjoint
union of $\overline\B$ with the  copies obtained
by antisymmetrizing in all the special punctures, a total of $n!$ properly
signed copies of $\overline\B$. Roughly speaking one cannot find in the
primitive domain two surfaces that just differ by the exchange of two
special punctures, such surfaces may appear in the boundary of the 
primitive domain. The primitive domain of a $\B$ space is not unique.

\section{Extended moduli spaces and the antibracket}

Let us first recall the definition of the antibracket for
ordinary moduli spaces of punctured surfaces.
Given two moduli spaces $\A_1$ and $\A_2$ their antibracket $\{ \A_1 , \A_2\}$
is defined as follows: pick any particular labeled puncture out of the
$n_1$ punctures of $\A_1$ and any
particular labeled puncture out of the $n_2$ punctures in $\A_2$
 and do twist sewing. In the resulting
space of surfaces, having a total of $n= (n_1-1) + (n_2-1)$ 
punctures we add over all inequivalent  ways
of splitting the $n$ punctures, now labelled from 1 to $n$, into
two groups of unordered punctures, one with $(n_1-1)$ punctures 
and the other one with $(n_2-1)$ punctures.
With this definition one readily verifies that [\senzwiebachtwo]  
$$\{\A_1 , \A_2\}= -\,(-)^{(\A_1+1)(\A_2+1)} \,\{\A_2,\A_1\}\,, \eqn\oneone$$
$$(-)^{(A_1+1)(A_3+1)} \Bigl\{ \{ \A_1 , \A_2\}\, 
 , \, \A_3\Bigr\}\,+\,
\hbox{cyclic} \, = 0\,\,,\eqn\twotwo$$
$$ \p\, \{\A_1 ,  \A_2\}  = \{ \p\A_1 ,  \A_2\} + (-)^{\A_1 +1} \{ \A_1 ,
\p\A_2 \} \; . \eqn\threethr$$

Let us now define the antibracket $\{\B_1 , \B_2\}'$ of two $\B$ spaces.
We include a prime because the final antibracket we are interested
in will have an extra sign factor.
We take $\B_1$ to be an $(n_1, \bar n_1)$ space and
$\B_2$ to be an $(n_2, \bar n_2)$ space. We first 
 compute the antibracket as if there
were only ordinary punctures. In all of the resulting surfaces
the total number of special punctures
is now $\bar n_1 + \bar n_2$. The labels of the special punctures
that arise from $\B_1$ are left unchanged, and the punctures that
arise from $\B_2$ are consecutively labelled  
$\bar n_1 +1 , \cdots , \bar n_1 + \bar n_2$.
 Take the complete set of labelled special punctures and consider
the inequivalent splittings of them 
into two groups $(i_1 \cdots i_{\bar n_1})$ and $(i_1 \cdots i_{\bar n_2})$  
with $\bar n_1$ and $\bar n_2$ elements respectively. Two splitting
are equivalent if the resulting groups 
are seen to contain the same labels, regardless of the order.
For each splitting label the ordered punctures in the $\B_1$ side of
the sewn surface using the
labels $(i_1 \cdots i_{\bar n_1})$ and label the ordered punctures in the 
$\B_2$ side of
the sewn surface using the
labels $(i_1 \cdots i_{\bar n_2})$. The sign factor assigned to this
splitting is the sign of the permutation necessary to  bring
$(i_1,  \cdots , i_{\bar n_1} , i_1, \cdots , i_{\bar n_2})$ to 
standard ascending
order. Add over all inequivalent splittings and that is the desired
antibracket. It follows from this definition that 
this antibracket will have the exchange property
$$\{\B_1 , \B_2\}' = -\,(-)^{(\B_1+1)(\B_2+1)+ \bar n_1 \bar n_2} 
\,\{\B_2,\B_1\}'\,, \eqn\newex$$
where the first part of the sign factor comes from the standard
antibracket, and the $\bar n_1 \bar n_2$ part arises because in
the exchanged antibracket the special
punctures of $\B_2$ should be labelled first.
The derivative property of the antibracket is not changed
$$ \p\, \{\B_1 ,  \B_2\}'  = \{ \p\B_1 ,  \B_2\}' + (-)^{\B_1 +1} \{ \B_1 ,
\p\B_2 \}' \; . \eqn\oldder$$
Finally, the Jacobi identity now takes the form
$$(-)^{(\B_1+1) (\B_3+1) + \bar n_1 \bar n_3} \{ \{ \B_1, B_2 \}' , \B_3 \}' 
 \, + \, \hbox{Cyclic}(1\to 2\to 3) = 0\,,\eqn\jacobini$$
where the extra sign factors with respect to the standard relation \jacobi\
appear due to the special punctures.

It is useful now to introduce the final version $\{\,\,\,,\,\,\}$ of the
antibracket of two $\B$ spaces by including a sign factor that will make
the identities take a more familiar form. We let
$$\{ \B_1 , \B_2 \} \equiv (-)^{{\bar n}_1 (1+ \B_2)} \{ \B_1 , \B_2
\}'\,.\eqn\fnalanb$$
With this redefinition the last three identities will read
$$\{\B_1 , \B_2\} = -\,(-)^{(\B_1+\bar n_1 +1)(\B_2+\bar n_2 + 1)} 
\,\{\B_2,\B_1\}\,, \eqn\newex$$
$$ \p\, \{\B_1 ,  \B_2\}  = \{ \p\B_1 ,  \B_2\} + (-)^{\B_1 +\bar n_1 +1} 
\{ \B_1 , \p\B_2 \} \; . \eqn\oldder$$
$$(-)^{(\B_1+\bar n_1 + 1) (\B_3+\bar n_3 + 1)} \{ \{ \B_1, B_2 \} , \B_3\} 
  \, + \, \hbox{Cyclic}(1\to 2\to 3) = 0\,,\eqn\jacobin$$
where we see that we have the standard identities if we assign to a
$\B$ space a grassmanality equal to the real dimension of the space plus
the number of special punctures.

\section{The $\K$ operator} 
On a surface that has only ordinary punctures the operator $\K$ has
been defined earlier [\senzwiebachtwo]; 
it inserts a special puncture, labelled ${\bar 1}$
throughout the
surface minus the unit disks around the ordinary punctures. In such
cases we have 
$$\eqalign{
\K\, (\,\{ \A_1 , \A_2\}\, )\, &=\, \{ \K\A_1 , \A_2\}
 \,+\,\{ \A_1 , \, \K \A_2 \} \,,\cr
[\,\partial\,,\,\K\,]\,&=\,-\,\{\, \V'_3\, , \,\, \}\,.\cr
}\eqn\oldprop$$
Now consider the case when we act on a surface which has one special puncture,
labelled ${\bar 1}$, 
in addition to several ordinary punctures. We will define the operation 
$\K$ to add a new special puncture, labelled ${\bar 2}$ and to insert it 
throughout the surface minus unit disks of the {\it ordinary} punctures, just 
ignoring the  special puncture ${\bar 1}$. To this result one must subtract
the same surfaces with the labels $\bar 1$ and $\bar 2$ exchanged, thus
producing a space of surfaces with two special punctures satisfying the
requisite antisymmetry property (see Figure 3).
The special punctures collide but divergences will
be avoided by antisymmetrization on the special punctures.
If the original surface has $\bar n$ special punctures $\K$ will insert
a puncture labelled ($\bar n +1$) throughout the surface minus the unit disks 
around the {\it ordinary} punctures. 
It will then subtract $\bar n$ copies of this surface
where the label of the new special puncture is exchanged with each of the
labels of the original special punctures, one at a time. This defines
$\K\B$.  At the level of primitive domains we have that 
$\overline{\K\B} = \K \overline\B$, where in the right hand side the
$\K$ simply adds the $(\bar n + 1)$ puncture.

\Figure{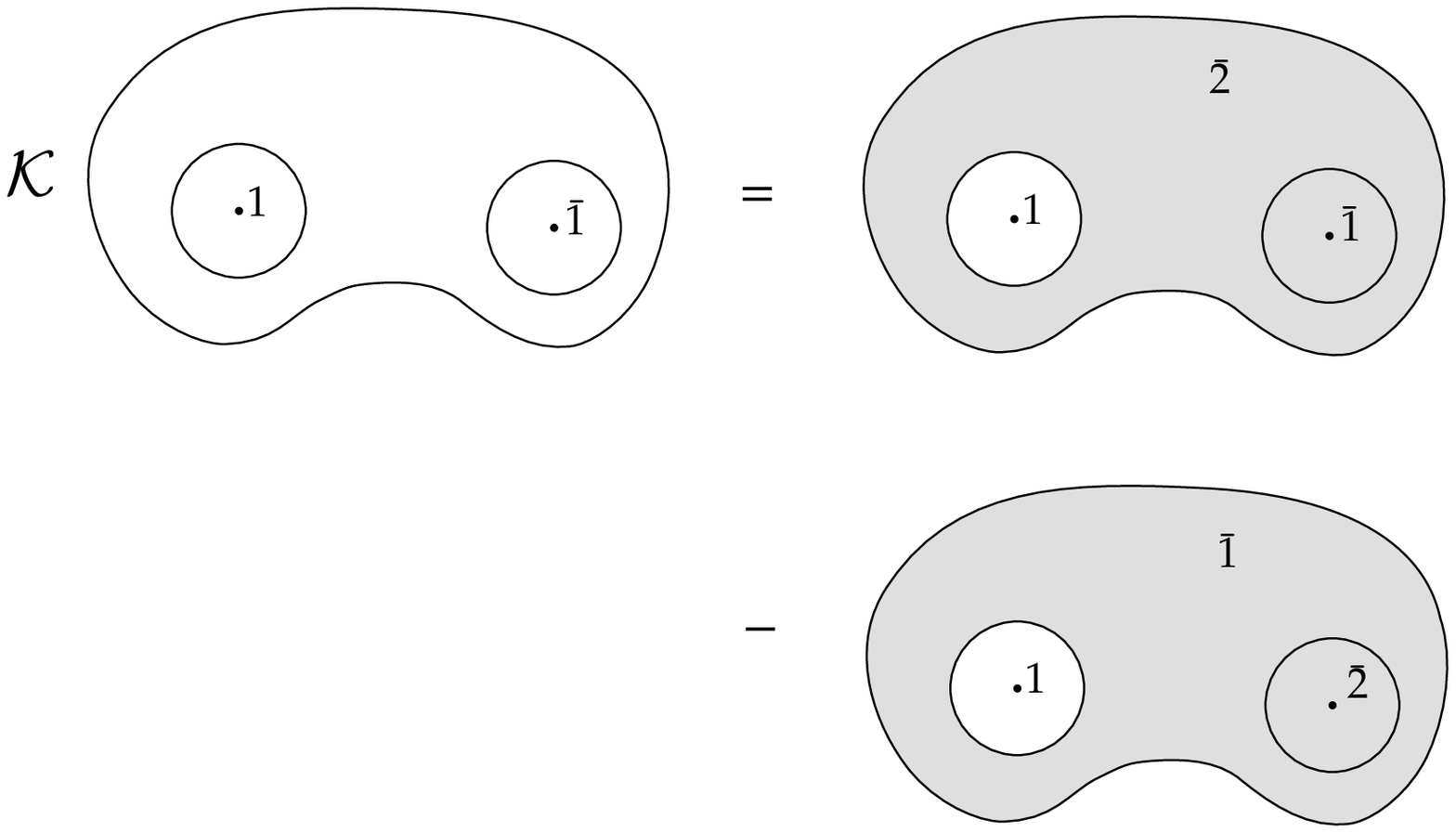 scaled 800}{Figure 3.~ This figure illustrates the
definition of the operator $\K$ when  it acts on a surface with one ordinary
puncture $1$ and one special puncture
$\bar 1$. The second special puncture $\bar 2$ is added throughout the surface minus
the unit disk around the ordinary puncture. Then one antisymmetrizes 
on the special punctures.}

We now claim
that the following three equations hold
$$\eqalign{
\K\, (\,\{ \B^{\bar n_1} , \B^{\bar n_2}\}'\, )\, &=\,(-)^{\bar n_2} 
\{ \K\B^{\bar n_1} ,\B^{\bar n_2}\}'
 \,+\,\{ \B^{\bar n_1} , \, \K\B^{\bar n_2} \}' \,,\cr
[\,\partial\,,\,\K\,]\, \B^{\bar n} &=\,-\,(-)^{\bar n}
\{\, \V'_3\, , \B^{\bar n} \}'\, = \,-\,(-)^{ \B}
\{\,\B^{\bar n}\,, \V'_3 \}'\,,\cr
\K ~\K & = 0\,\,. \cr} \eqn\knewd$$
Let us begin with the last equation. It holds acting on
a space with no special punctures because the two special punctures
end up integrated over the same region, and the antisymmetry condition
will make them cancel out. This is the way our computation of curvature
in the last section worked out. On a space that already has some special
punctures the regions of integration still agree (given that all special
punctures are ignored when inserting the new puncture) and the configurations
where the two new special punctures have the labels $\bar n + 1$ and 
$\bar n + 2$ clearly cancel. All other configurations will also cancel
out in pairs.

The first equation is also simple to understand. Consider the configuration
where in the left hand side $\K$ inserts the puncture labelled
$(\bar n_1 + \bar n_2 + 1)$ on the
sewn surface. If the puncture lies on the $\B^{\bar n_2}$ surface then the
labelling of the punctures is consistent with the antibracket of the
second term in the right hand side and no sign
factor is required. If it lies on the $\B^{\bar n_1}$ surface, 
to compare with the labelling implied by the first term in the 
right hand side we
have to relabel all the punctures in the second surface, a total of ${\bar n_2}$
exchanges giving rise to the sign factor. 

In the second equation the sign factor that appears in addition
to the sign factor in \oldprop\ also arises for
similar reasons. In the left hand side, 
the puncture that ends on the boundary of
a disk is  the $\bar n + 1$ puncture. On the other hand,
in the right hand side it is the first special puncture that ends there,
since $\V'_3$ appears first in the antibracket. The rearrangement of labels
requires $\bar n$ exchanges, thus the sign factor.

Finally, using the unprimed antibracket (\fnalanb) the
equations listed in \knewd\ become
$$\eqalign{
\K\, (\,\{ \B^{\bar n_1} , \B^{\bar n_2}\}\, )\, &=\,(-)^{\B_2 + \bar n_2 +1} 
\{ \K\B^{\bar n_1} ,\B^{\bar n_2}\}
 \,+\,\{ \B^{\bar n_1} , \, \K\B^{\bar n_2} \} \,,\cr
[\,\partial\,,\,\K\,]\, \B^{\bar n} &=\, (-)^{\B + \bar n}
\{\, \V'_3\, , \B^{\bar n} \}\, \,,\cr
\K ~\K & = 0\,\,. \cr} \eqn\knewd$$

\section{$\I$ operator and $\{ \I \, , \, \K \}$.}~ 
The operator $\I$  acting on a moduli space
whose surfaces have only ordinary punctures,  will pick a fixed label puncture
and convert it into a special puncture, labelled ${\bar 1}$. 
Which puncture is chosen is irrelevant
since the space is symmetric under any exchange of ordinary punctures (the
remaining ordinary punctures may be relabelled with consecutive labels, if
necessary). This operation was not defined explicitly earlier, 
underbars were used to indicate spaces of ordinary punctures 
that happened to get a special state inserted in one of the punctures. 
It is useful, however, to bring this operation into the open.
The following identities are simple consequences of the above definition
$$\eqalign{
\I\, (\,\{ \A_1 , \A_2\}\, )\, &=\, \{ \I\A_1 , \A_2\}
 \,+\,\{ \A_1 , \, \I \A_2 \} \,,\cr
[\,\partial\,,\,\I\,]\,&=\,0\, \,. \cr}\eqn\orde$$
On a $\B$ space  with $\bar n$ special punctures the operator $\I$
will change one of the ordinary punctures into a new
special puncture and will label
it $\bar n +1$. Then it will subtract ${\bar n}$ copies, exchanging the
label of the new special puncture with each of the ${\bar n}$ labels of
the original special punctures (see Figure 4). This defines $\I\B$.
At the level of fundamental domains we have
$\overline{\I\B} = \I\overline\B$, where the $\I$ in the right hand 
side just converts one puncture and labels it $(\bar n +1)$. 
The operator $\I$ is rather similar to $\K$, except
that the dimensionality of the space of surfaces is not changed by the
action of $\I$, nor are new collisions of punctures induced. We claim that
the following identities hold 
$$\eqalign{
\I\, \{ \B^{\bar n_1} , \B^{\bar n_2}\}\,  &=\,(-)^{\B_2 + \bar n_2 +1} 
\{ \I\B^{\bar n_1} ,\B^{\bar n_2}\}
 \,+\,\{ \B^{\bar n_1} , \, \I\B^{\bar n_2} \} \,,\cr
[\,\partial\,,\,\I\,]\, &=\,0 \,,\cr
\I ~\I & = 0\,\,. \cr}\eqn\iinst$$
The first equation here follows directly from the first equation of \orde\
with the sign factors identical to those appearing in \knewd. The second
identity needs no explanation.
The last identity follows because after acting twice with $\I$, a space will
have two ordinary punctures turned special, and the antisymmetrization
on the special punctures will give zero due to the original symmetry 
under exchange of ordinary punctures.

\Figure{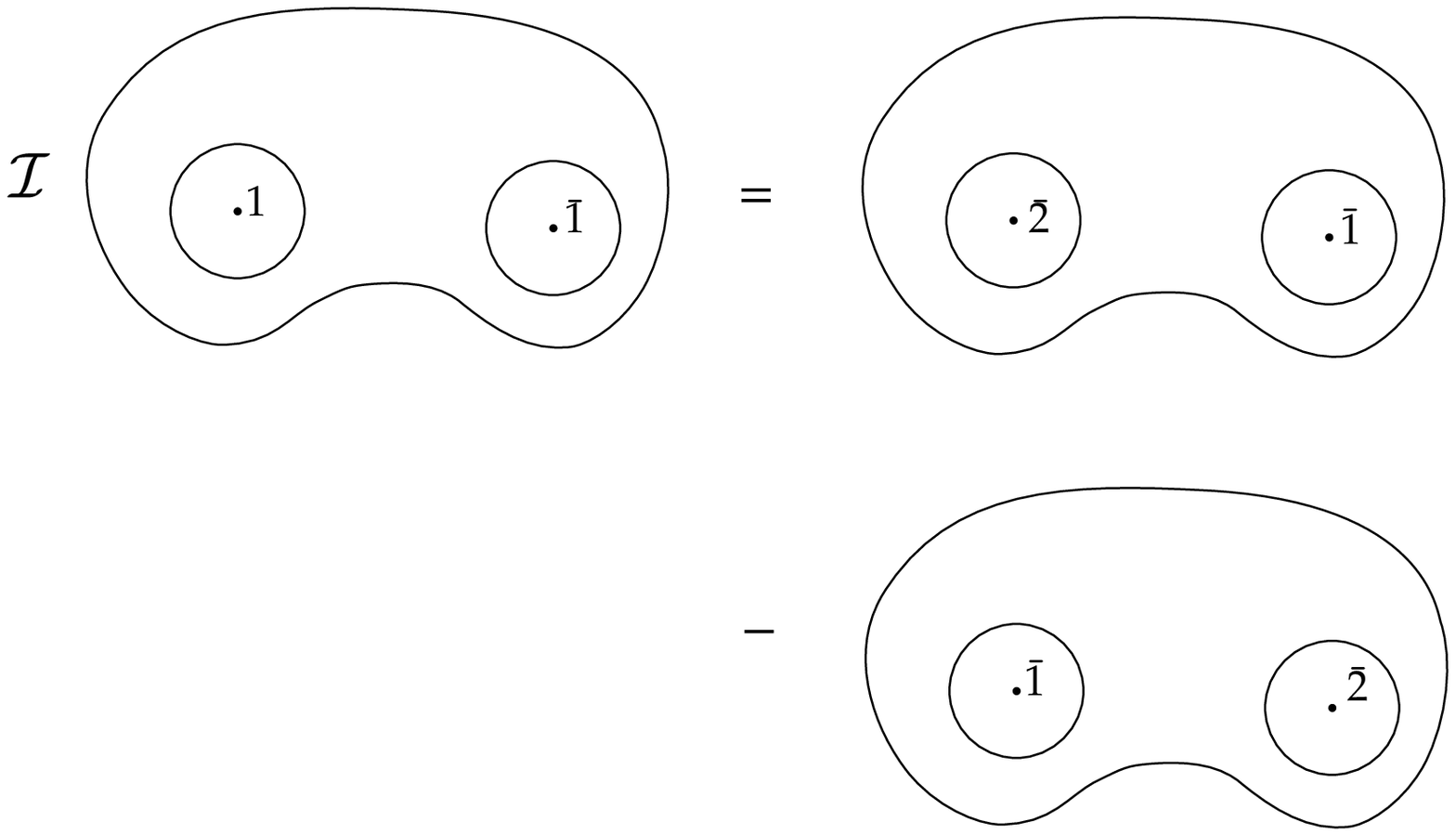 scaled 800}{Figure 4.~ This figure illustrates the
definition of the operator
$\I$ when  it acts on a surface with one ordinary puncture $1$ and one special
puncture $\bar 1$. The ordinary puncture is made special and labelled 
$\bar 2$. Then one antisymmetrizes on the special punctures.}

There is one important relation we discuss now. We claim that
$$ \K~ \I + \I ~\K = \{ \,\,\,  , \T^2_1\,\,\}\,\,,\eqn\import$$   
where $\T^2_1$ is the $\B$ space with two special punctures introduced
in section 4.3. 

As a preliminary point note that $\{ \,\,,\,\T^2_1 \} = \{ \,\,, \T^2_1 \}'$,
regardless of the moduli space appearing in the first slot of the 
antibracket. This follows from Eqn.\fnalanb\ given that the
dimension of $\T^2_1$ is one. We can therefore discuss \import\ using
the primed antibracket.

As usual it is convenient to first discuss this
equation acting on ordinary moduli spaces. It is in fact sufficient to discuss
the case when we act on a single surface having a single ordinary puncture
and an associated canonical disk. This is illustrated in Fig.5.

\Figure{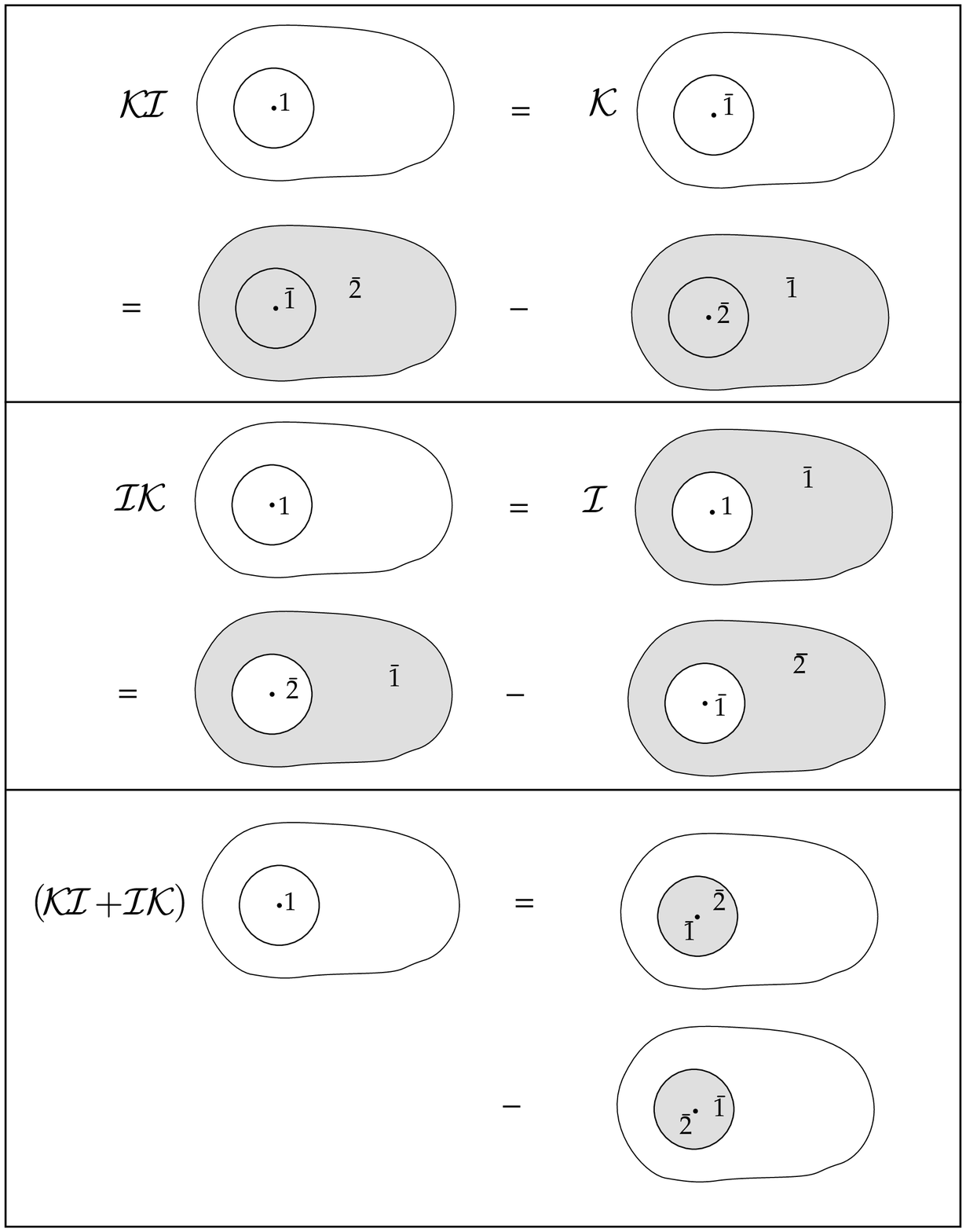 scaled 800}{Figure 5.~We illustrate the computation of $(\K\I
+\I\K)$ acting on a surface with one ordinary puncture. The end result shows 
one of the special punctures sitting at the ordinary puncture, while the
other one moves throughout the unit disk (with the usual antisymmetrization).}

The term $\K\I$
gives the surfaces where  the original puncture is labeled $\bar 1$ and the
second puncture $\bar 2$ moves throughout the surface, minus the surfaces with
the labels exchanged. The term $\I\K$ gives surfaces where the original
puncture is labelled $\bar 2$ and a puncture $\bar 1$ moves on the surface
minus the unit disk, minus the surfaces with the labels exchanged. It follows
that the sum of terms gives surfaces where the original puncture is labelled
$\bar 1$ and the puncture
$\bar 2$ moves {\it within} the unit disk, minus the surfaces with the labels
exchanged. These moduli spaces have an orientation which is
that induced by the standard orientation of the disk where the puncture
moves. Up to a sign this is precisely what the space $\T^2_1$ would do 
upon sewing via the antibracket. 

Let us now justify the sign in the equation. It follows
from the definition below \onedim\ and the relation \intdefd\ that
the surfaces with puncture $\bar 1$ sitting and puncture $\bar 2$ moving
radially away appear in $\T^2_1$ with a {\it minus} sign. Call the radial vector
$\partial/\partial r$. The twist sewing of the antibracket
involves the rotation angle $\theta$ with
the associated vector $\partial/\partial \theta$. The orientation of the
right hand side is defined by the ordered basis $[\partial/\partial \theta,
\partial/\partial r]$, which is opposite to the standard orientation of
the disk. This induces a new minus sign, and we conclude that the overall
sign must be a plus, as indicated in the equation. The extension to the case
when we act on $\B$ spaces is now immediate. Since the space $\T^2_1$ appears
as the second entry in the antibracket, no new signs will be induced, the
new special punctures will have the same labels in the left hand side and
in the right hand side.

\section{The $\M$ operator.}~  We now define $\M \equiv \K - \I$. This operator 
measures the difference between
adding a special puncture and making an ordinary puncture special.
We have given it a special name since it appears often enough.
It follows from the  identities given in \knewd\ and \iinst\ that 
$$\eqalign{
\M\, \{ \B^{\bar n_1} , \B^{\bar n_2}\}\,  &=\,(-)^{\B_2 + \bar n_2 +1} 
\{ \M\B^{\bar n_1} ,\B^{\bar n_2}\}
 \,+\,\{ \B^{\bar n_1} , \, \M\B^{\bar n_2} \} \,,\cr
[\,\partial\,,\,\M\,]\,\B &=\,(-)^{\B + \bar n}
\{\,\V'_3\,, \B\,  \}\,,\cr
\M ~\M & = - \, \{ \,\,\,  , \T^2_1\,\,\}\, .\cr}\eqn\mopiden$$
In deriving the last equation one makes
use of \import.

It will be useful to bring in the operator $\delta_\V$ defined in
\anuop. It is a short calculation to prove that  
$$\eqalign{
[\, \,\delta_\V ,\K\, ]\B &= (-)^{\B+ \bar n}
\bigl\{ \, \V'_3 + \K\V \, ,  \B\, \bigr\}\,, \cr 
[\,\,\delta_\V ,\I\, ]\B &= \, (-)^{\B + \bar n} \bigl\{  \,\I\V\,,   \B\,
\bigr\} \,, \cr
[\,\delta_\V ,\M\, ]\B &= \,(-)^{\B +\bar n}\bigl\{  \, \V'_3 + \M\V\,, \B\, 
\bigr\}\,. \cr} \eqn\fcomb$$

\section{From Riemann surfaces to functions}  

We recall from Ref.[\senzwiebachtwo] that there is a homomorphism
from the BV algebra of Riemann surfaces to the BV algebra of functions
in $\H$. One defines, for ordinary moduli spaces of surfaces
$$f(\A) \,\equiv {1\over n!}\,
\int_{\A}\bra{\Omega}\Psi\rangle_1\cdots
\ket{\Psi}_n\,\, , \quad n\geq 1\, ,\eqn\homfun$$
and shows, using \antibnn\ and Eqn.(3.12) of Ref.[\senzwiebachtwo] that
$$f \bigl( \{ \A_1 , \A_2 \} \bigr) = - \{\, f(\A_1) , f(\A_2) \}\, .\eqn\hone$$
For the action of the BRST operator one finds
$$\{ \,Q , f (\A )\} = \, - \,f\bigl( \p \A\bigr)\,\,,
\, \quad Q = \half
\,\bra{\omega_{12}}
\widehat Q^{(2)}\ket{\Psi}_1\ket{\Psi}_2\,. \eqn\brstact$$
We need to extend these results to the case when the moduli space of
surfaces has special punctures. For moduli spaces with 
$n$ ordinary punctures and one or two special
punctures respectively, we define 
$$\eqalign{
f_{\O_1}(\B) \,&\equiv
{1\over n!} \int_{\B} \bra{\Omega\,}
\Psi\rangle_1\cdots\ket{\Psi}_{n}\ket{\O_1}_{\bar 1}\,, \cr
f_{\O_1\O_2}(\B) \,&\equiv {1\over n!} \int_{\B} \bra{\Omega\,}
\Psi\rangle_1\cdots\ket{\Psi}_{n}\ket{\O_1}_{\bar 1}\ket{\O_2}_{\bar 2}\,.\cr}
\eqn\twopu$$
Antibrackets between ordinary moduli spaces and special moduli spaces 
are readily found to map as
$$\eqalign{
f_{\O_1} \bigl( \{ \A , \B \} \bigr) &=
- \{\, f(\A) , f_{\O_1}(\B) \}\,,\cr
f_{\O_1\O_2} \bigl( \{ \A , \B \} \bigr) &=
- \{\, f(\A) , f_{\O_1\O_2} (\B) \}\,. \cr}\eqn\simpanti$$
Note that given \fnalanb\ there is no difference between $\{ \A , \B\}'$
and $\{ \A , \B \}$. We can thus justify \simpanti\ using the primed
antibracket, and then simply use the unprimed one. 

The antibracket between two moduli spaces each of one
special puncture is given by
$$f_{ij}\Bigl( \{ \B , \B'\}' \Bigr) = (-)^{1+ i (\B'+1)} \Bigl[
\bigl\{ f_i (\B) , f_j (\B') \bigr\}  
+ (-)^{(i+1)(j+1) + \B'(i+j) } \bigl\{ f_j (\B) , f_i (\B') \bigr\} \Bigr]\,.
\eqn\ovd$$
Here we use subscripts $i$ and $j$ instead of $\O_i$ and $\O_j$, for the
sake of brevity. The right hand side shows two terms, which arise because
the antibracket in the left hand side involves an antisymmetrization on
the special punctures (see sect.4). The sign factors have been given in
all generality.  If the states are both grassmann even, 
the above equation gives
$$f_{\mu\nu}\Bigl( \{ \B , \B'\}' \Bigr) = -
\bigl\{ f_\mu (\B) , f_\nu (\B') \bigr\}  
+ \bigl\{ f_\nu (\B) , f_\mu (\B') \bigr\}\,.   
\eqn\strom$$
If, in addition, $\B=\B'$, and dim~$(\B)$ is odd we find (see \fnalanb)
$$f_{\mu\nu}\Bigl( \{ \B , \B\} \Bigr) = - 2
\bigl\{ f_\mu (\B) , f_\nu (\B) \bigr\} \,, 
\eqn\stromp$$
where we now use the unprimed antibracket in the left hand side.
This equation holds, by virtue of
\strom, when the space
$\B$ is actually a sum of spaces, all with one special puncture and all
odd-dimensional.

Let us now consider the homomorphism identities relevant to the operator $\I$.
The linear hamiltonian $B^{(2)}_\O$ (formerly called $U^{(2)}_\O$) is the
hamiltonian  required to produce insertions of special states on ordinary
punctures 
$$B^{(2)}_{\O}\equiv \bra{\omega_{12}\,}\O\rangle_1\ket{\Psi}_2\,\,.\eqn\lham$$
One readily finds that the antibracket of two such hamiltonians is a
constant hamiltonian
$$  \{ B^{(2)}_{\O_1},B^{(2)}_{\O_2}\} = 
\bra{\omega_{12}\,}\O_1\rangle_1\ket{\O_2}_2 . \eqn\lhambr$$
Inserting states on moduli spaces of surfaces with ordinary punctures
is straightforward
$$f_{\O} (\I\A) = \,\{\, f(\A)\, , \,B^{(2)}_{\O}\,\}\,  \,.\eqn\iord$$
To insert a state on a moduli space of surfaces that already has one
special puncture, we recall that the operator $\I$ will convert an
ordinary puncture into a special one and then will antisymmetrize
on the special punctures. One finds
$$f_{ij} (\I \B) = \{ f_i (\B) \, , B_{j}^{(2)} \}
+ (-)^{i (\B + 1) } \{ B_{i}^{(2)}\, ,f_j (\B) \} \,,\eqn\ispec$$
where again, for brevity, we  $i$ and $j$ instead of $\O_i$ and $\O_j$.
For grassmann even states the above simply becomes
$$f_{\mu\nu} (\I \B) = \{ f_\mu (\B) \, , B_{\nu}^{(2)} \}
+  \{ B_{\mu}^{(2)}\, ,f_\nu (\B) \} \,.\eqn\ispecx$$

BRST action on ordinary moduli spaces is familiar (see, for example 
[\senzwiebachtwo])  
$$\{ \,Q \,,\, f(\A)\} = \, - \, f(\p\A)\, \, .\eqn\brstms$$
If we act on a $\B$ space with one special puncture we find
$$\{ \,Q \,,\, f_\O (\B)\} = \, - \, f_\O(\p\B)\, + \,
  (-)^\B  f_{Q\O}(\B) \,,\eqn\newactb$$
by doing a computation virtually identical to that used to establish
Eqn.(2.41) of Ref.[\bergmanzwiebach]. For $\B$ spaces with two or more
punctures there is only little difference
$$\{ \,Q \,,\, f_{\O_i\O_j} (\B)\} = \, - \, f_{\O_i\O_j}(\p\B)\, + \,
  (-)^\B [ f_{(Q\O_i)\O_j} + (-)^if_{\O_iQ\O_j} ]\, (\B) \,,\eqn\newactb$$
For our case of interest we will look at $f_{\mu\nu} = f_{\wh\O_\mu\wh\O_\nu}$
and since both $\wh\O_\mu$ and $\wh\O_\nu$ are annihilated by $Q$, we find
$$\eqalign{ \{ S , f_{\mu\nu}(\B) \} &= 
\{\,  Q + f(\V)\, , f_{\mu\nu} (\B) \} \cr 
&=  - f_{\mu\nu} \bigl(  \partial \B + \{ \V\,,\, \B \}  \bigr) \,,\cr
&= - f_{\mu\nu} \bigl(  \delta_\V \B  \bigr)\,,\cr}\eqn\newde$$
where use was made of \simpanti.

With the canonical connection, covariant derivatives of
hamiltonians arising from  ordinary moduli spaces are readily computed
[\senzwiebachtwo]
$$D_\mu  \, f(\A ) = f_\mu\, ( \K \A )\,. \eqn\kcov$$
For moduli spaces with one special puncture filled with a marginal operator
($\wh\O_\mu, \wh\O_\nu, \cdots$), antisymmetric covariant
derivatives take a simple form. Consider the covariant derivative
$D_\mu f_\nu (\B)$. It follows from the first equation in \twopu\ that
there are two contributions, one from the derivative of the surface state
and the other from the derivative of $\ket{\wh\O_\nu}$. We find
$$D_\mu \, f_{\nu}(\B) =  
f_{\nu\mu}(\widetilde\K \B)
+f_{ D_\mu \ket{\wh\O_\nu} } (\B)\,, \eqn\eqntthree$$
where $\widetilde\K$ denotes
operation of adding a puncture, labelled $\bar 2$, where $\wh\O_\mu$
is inserted,  avoiding all coordinate disks {\it including}
that of the special puncture. Note that $\widetilde\K$ performs
no antisymmetrization. 
Forming the antisymmetric combination
we find
$$D_\mu \, f_{\nu}(\B)- D_\nu \, f_{\mu}(\B) =  
f_{\nu\mu - \mu\nu}(\widetilde\K \B)
+f_{ D_\mu \ket{\wh\O_\nu} -  D_\nu \ket{\wh\O_\mu} } (\B) \,.\eqn\terminus$$
Using \gtor, we recognize that the second term in the 
right hand side is  making an insertion throughout the disk associated with
the special puncture. In fact, the signs work out correctly (compare
\gtor\ with \conder) so that the two terms in the right hand side combine
to make insertions throughout the surface avoiding only the disks associated
with ordinary punctures.  
This is precisely the operation $\K$, which includes antisymmetrization.
We have therefore shown that 
$$D_\mu \, f_{\nu}(\B)- D_\nu \, f_{\mu}(\B) =  
- f_{\mu\nu}(\K \B)\,. \eqn\hardin$$

\chapter{Consistency Conditions for Moduli Spaces}

In this section we will examine explicitly the consistency conditions
derived in  section 3.  They read
$$  D_\mu B_\nu -D_\nu B_\mu +\{ B_\mu  , B_\nu \}\, +\,  \,R_{\mu\nu}=  
\bigl\{ \,\,  S\, , \, B_{\mu\nu} \, \bigr\} \,. \eqn\eightt$$
Since the above covariant derivatives are taken with respect to
the canonical connection $\Gamma_\mu$ and this connection has
vanishing curvature $R_{\mu\nu}\,  (\,\Gamma\, ) = 0$, the
equation to be discussed is 
$$  D_\mu B_\nu -D_\nu B_\mu +\{ B_\mu  , B_\nu \}\, =  
\bigl\{ \,\,  S\, , \, B_{\mu\nu} \, \bigr\}\,.  \eqn\disc$$
The aim of the present section is to show that the hamiltonian $B_{\mu\nu}$
is the function associated to some moduli spaces of surfaces with two 
special punctures. We will use our analysis of the previous section to
derive the recursion relations that must be satisfied by the moduli spaces
that define the hamiltonian $B_{\mu\nu}$.

We therefore take  $B_{\mu\nu}$ to be a hamiltonian of the form 
$$B_{\mu\nu} = -f_{\mu\nu} \,( \B^2) = -\int_{\B^2}\bra{\Omega_{\bar 1\bar 2}}
\O_\mu\rangle_{\bar 1} \ket{\O_\mu}_{\bar 2}  \,,  \eqn\nota$$
where $\B^2$ is a sum of moduli spaces of surfaces with two special
punctures. The spaces in $\B^2$ are
antisymmetric under the exchange of labels of the special punctures,
$\bra{\Omega_{\bar 1\bar 2}} = -\bra{\Omega_{\bar 2\bar 1}}$, and this
explains why $B_{\mu\nu} = -B_{\nu\mu}$. 

The right hand side of \disc\ is readily written in terms of moduli spaces 
$$\bigl\{ \,\,  S\, , \, B_{\mu\nu} \, \bigr\} =
-\bigl\{ \,\,S\, , \, f_{\mu\nu} (\B^2)  \, \bigr\}
= \, f_{\mu\nu} \bigl( \delta_\V  \B^2 \bigr) \,,  \eqn\hober$$
where use was made of Eqn.\newde. 
Let us now compute $(D_\mu B_\nu  - D_\nu B_\mu)$. Since
$B_\nu = B^{(2)}_\nu - f_{\nu}(\B^1)$, we must consider
$$D_\mu B_\nu  - D_\nu B_\mu = (D_\mu B^{(2)}_\nu - D_\nu B^{(2)}_\mu)
- ( D_\mu \, f_{\nu}(\B^1)- D_\nu \, f_{\mu}(\B^1) )  \eqn\consi$$
The first parenthesis is readily computed
$$D_\mu B^{(2)}_\nu - D_\nu B^{(2)}_\mu 
= B^{(2)}_{ D_\mu \ket{\O_\nu} -  D_\nu \ket{\O_\mu}} 
= f_{\mu\nu} (\T_1^2) \,,\eqn\ttwo$$
where we made use of \bbb. The second parenthesis was computed in \hardin, 
so all together we find 
$$D_\mu B_\nu -D_\nu B_\mu =f_{\mu\nu}( \T_1^2 +\K\B^1)\,. \eqn\trantor$$
On the other hand 
$$\eqalign{
\{ B_\mu , B_\nu \} &= \,\,\,\,\{ B_\mu^{(2)} , B_\nu^{(2)} \} \cr
&\quad  -\{ f_\mu (\B^1) , B_\nu^{(2)}\}  -\{ B_\mu^{(2)} , f_\nu (\B^1) \}\cr
&\quad  + \{ f_\mu(\B^1) \,, \, f_\nu (\B^1)\}  \cr} \eqn\salvor$$
The first line in the right hand side vanishes by virtue of \lhambr,
the second line is simplified using \ispec\ and the third
line  using
\stromp. We thus find
$$\{ B_\mu , B_\nu \} = - f_{\mu\nu}  \bigl(\, \I \B^1 
+ \, \half \,\{ \B^1 \,, \, \B^1 \}\, \bigr) \,. \eqn\seldon$$
It is now possible to put together Eqns.\hober,\trantor, and \seldon\ to find
$$   f_{\mu\nu}\Bigl(\,\T^2_1 +  \M \,\B^1  -\, \half\,
 \{ \B^1 \,,\, \B^1 \} \Bigr) =
 f_{\mu\nu} \bigl( \delta_\V  \B^2 \bigr) \,. \eqn\mallow$$
This equation will be satisfied if 
$$  \delta_\V  \B^2  = \T^2_1 +  \M \,\B^1  -\, \half\,
 \{ \B^1 \,,\, \B^1 \} \,.\eqn\gmallow$$
Note that by construction the surfaces  in the right hand
side of the equation 
are all $\B^2$ spaces, namely moduli spaces of surfaces with two
special punctures,  antisymmetric under the exchange of those punctures.
The antisymmetry of the hamiltonian $B_{\mu\nu}$ implied that $\B^2$ had
to be a $\B$ space, and the action of $\delta_\V$ does not change the fact
that it is a $\B$ space. Thus the above equation is an equation relating
$\B$ spaces. 

Let us now verify the consistency of \gmallow.
We should be able to check that $\delta_\V\,\delta_\V  \B^2 =0$. 
We calculate the action of $\delta_\V$
on each term of the right hand side of \gmallow\ and find
$$\eqalign{
\delta_\V\,\T^2_1  &= \partial \T^2_1  + \{\V ,  \T^2_1   \} = \I\V'_3 +  
\{\V ,  \T^2_1   \} \,, \cr &\cr
\delta_\V \M \B^1 &= [\delta_\V , \M ] \B^1 + \M \delta_\V \B^1  \cr
& =  \{ \V'_3 + \M \V\,, \B^1   \} + \M ( \V'_3 + \M \V )\,\cr
&= \{\V'_3 + \M \V \,,  \B^1  \} - \I \V'_3  - \{ \V , \T^2_1 \}\,, \cr & \cr
\delta_\V \bigl( -\, \half\,
 \{ \B^1 \,,\, \B^1 \} \bigr) &= - \{ \delta_\V \B^1 , \B^1 \} = 
- \{\V'_3 + \M \V \,,  \B^1  \}\,.\cr} \, \eqn\ufr$$
Back in \gmallow\ we see that all terms cancel out, and thus the recursion
relations are consistent. To derive the first equation we used \boundtau,
and for the second  we used \fcomb,~\recbb,~\newex, and \mopiden. 
We must now show that we can actually define the $\B^2$ spaces of the left
hand side of \gmallow. This is the subject of the next section.

\chapter{Recursive construction of $\B^2$ spaces}
\medskip
\nobreak
In order to construct the $\B^2$ spaces we must write the consistency
condition \gmallow\ as a recursion relation. To this end we rewrite this
equation as
$$  \partial  \B^2  = \T^2_1 +  \M \,\B^1  -\, \half\,
 \{ \B^1 \,,\, \B^1 \}  - \{ \V \, , \B^2 \} \,. \eqn\gmallows$$
It is worthwhile noticing that
the consistency check $\delta_\V^2 \B^2=0$ we  performed in the previous
section is equivalent to checking
that $\partial\partial \B^2= 0$. This establishes that the right hand side of
\gmallows\ indeed has zero boundary, a necessary, though not sufficient
condition for it to be a boundary. Showing it is a boundary is the purpose
of the present section. Our analysis will be very explicit for the 
$\B^2$ spaces with the lowest number of ordinary punctures. We then
outline a construction for the $\B^2$ spaces with higher number of ordinary
punctures. 

It is straightforward to see that \gmallows\ is a set of recursion relations.
Equation \gmallows\ gives an equation for each number of ordinary punctures. 
All $\B^2$ spaces have two special punctures, and we will define them
recursively on the number of ordinary punctures.
For any chosen number
$k$ of ordinary punctures $\partial\B^2_k$ involves either known spaces,
($\B^1$ or $\V$ spaces) or $\B^2_{k'}$ spaces with $k'<k$, given that 
$\V$ has at least
three ordinary punctures and thus the antibracket adds at least two
ordinary punctures. The lowest number of ordinary punctures appearing
in the right hand side is one. Therefore we will have
$$\B^2 \equiv  \B^2_1 + \B^2_2 + \B^2_3 + \cdots \, \, . \eqn\whsp$$
The first two equations that follow from \gmallows\ are
$$\eqalign{\partial \B^2_1 
 & =  \T^2_1 - \I\B^1_2\,,  \cr 
\partial \B^2_2 &=   \K \,\B^1_2 - \I \B^1_3  -\, \half\,
 \{ \B^1_2 \,,\, \B^1_2 \}  - \{ \V_3 \, , \B^2_1 \}\, .\cr } \eqn\frsttw$$
Let us first discuss explicitly the first equation. 

\section{Construction of $\B^2_1$}

We  want to construct the space $\B^2_1$ that satisfies $\partial \B^2_1 
=\T^2_1 -\I\B^1_2\,$.
We recall from \boundtau\ and \nytur\ that
$$\partial\T^2_1 = \I \V'_3\,,\quad\quad\quad 
\partial (\I\B^1_2) =\I\partial\B^1_2=\I (\V'_3-\I\V_3) = \I\V'_3\,.\eqn\hjj$$
These two relations explain that $\partial^2\B^2_1=0$, but in addition they
help us understand the geometrical picture, as is shown in Fig.6. 
It is useful to
consider the primitive domains of the moduli spaces $\T^2_1$ and $\I\B^1_2$:

$\bullet$ The primitive domain
$\overline{\T^2_1}$ is the set of surfaces $I S(t)$ ($I$ is the
same as $\I$ except that it does not include antisymmetrization on
the special punctures)
with $t\in [0,1]$, and thus extending from the 
sphere $I S(0)$ up to
$I\V'_3$. Let $\overline{\T^2_1}\,[t_0]$ denote the particular
three-punctured sphere obtained for $t=t_0$.

$\bullet$ The primitive domain $\overline{\I\B^1_2}$ is the set of surfaces 
interpolating from $II\V_3$ up to $I\V'_3$, 
a set that can also be
parametrized by $t\in [0,1]$. Similarly,
let $\overline{\I\B^1_2}\,[t_0]$ denote the particular
three-punctured sphere obtained for $t=t_0$. 

The two primitive spaces meet at 
the surface $I\V'_3$. Moreover
$$\partial (\overline{\T^2_1} - \overline{\I\B^1_2} ) = I S(0) -
II\V_3\,\,,\eqn\partb$$
and thus the primitive domain of the space in the right hand side
of the recursion relation extends from the totally
symmetric three punctured sphere
$II\V_3$ to the sphere $IS(0)$, which is symmetric in the two special
punctures it has (the two punctures actually coincide).
Recall that all surfaces involved here are three punctured
spheres and therefore they only differ by the choice
of local coordinates at their punctures. 

$\bullet$ Let $\Sigma (s, t_0)$ with $s\in [0,1]$
denote the set of spheres interpolating from 
$\Sigma (0, t_0)=\overline{\I\B^1_2}\,[t_0]$ 
up to $\Sigma (1, t_0)=\overline{\T^2_1}\,[t_0]$. All interpolations
are done canonically (as in Ref.[\senzwiebach] Sect.3.4).

\Figure{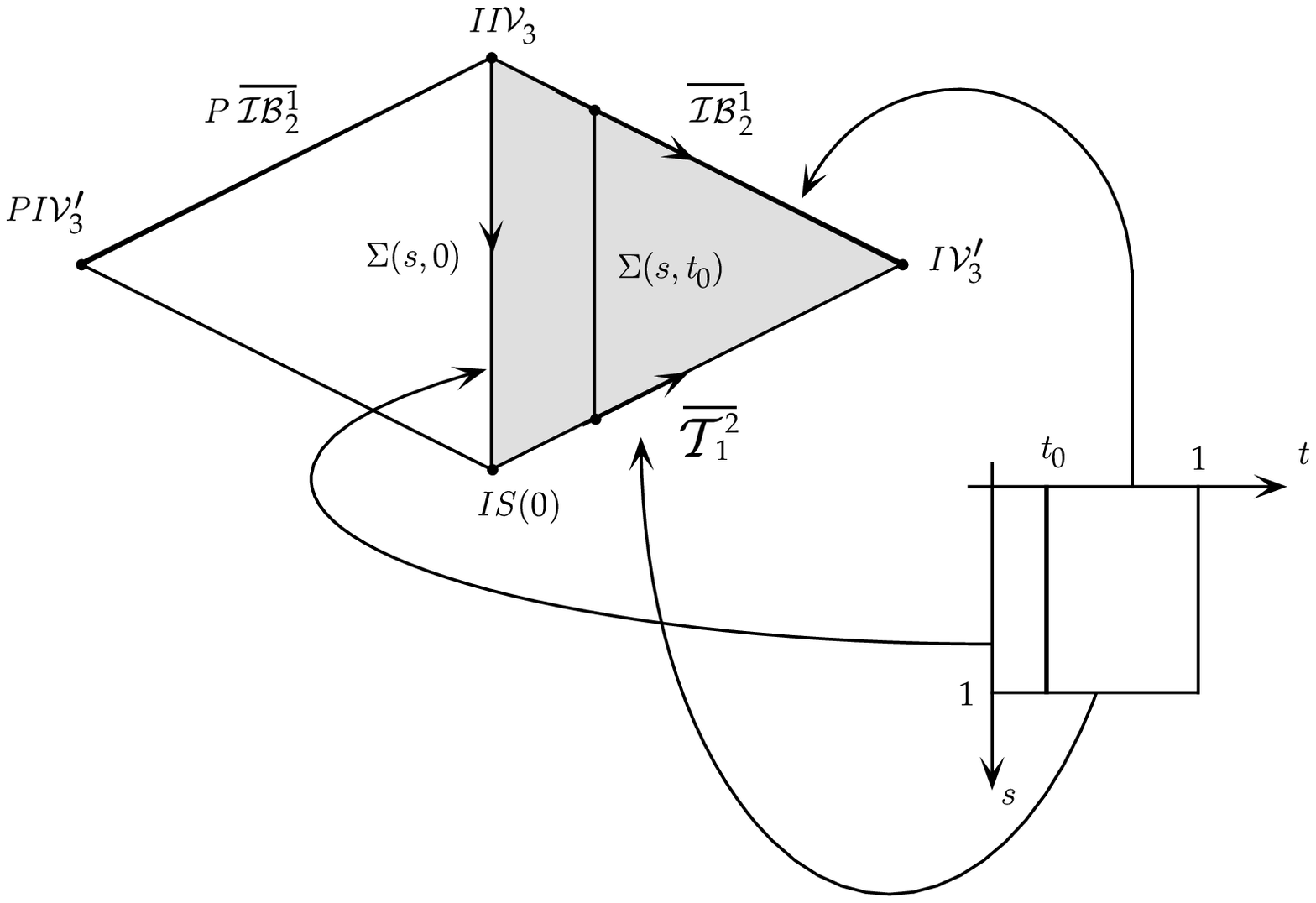 scaled 750}{Figure 6.~ The rhombus shaped region 
represents the desired space $\B^2_1$.
Its primitive domain is shown shaded. The boundary of the primitive domain
is $\overline{\T^2_1}- \overline{\I\B^1_2}$ plus the set of surfaces
$\Sigma (s,0)$ interpolating symmetrically from $II\V_3$ up to $IS(0)$.}

We now define the object $\overline {\B^2_1}$
$$\overline {\B^2_1} \equiv \{ \Sigma (s,t)\, \bigl| s, t\in [0,1] \} \,,
\eqn\thesp$$
with orientation defined by the ordered basis
$[{\partial\over \partial s}, {\partial\over \partial t}]$. It follows from
this definition that
$$\partial \overline {\B^2_1}  = \overline{\T^2_1} - \overline{\I\B^1_2} \,
+ \Sigma (s, 0) \,, \eqn\djeri$$
where the last term denotes the surfaces interpolating canonically
from the symmetric
vertex $II\V_3$ up to the symmetric vertex 
$IS(0)$. It follows from the nature of canonical interpolation
that all surfaces in $\Sigma (s, 0)$
are symmetric under the exchange of the two special punctures.
Note that there is no contribution to the boundary from $\Sigma (s,1)$
since it is an interpolation between identical surfaces
$({I\V'_3})$. We claim that, as the notation suggests, 
$\overline {\B^2_1}$ is the fundamental domain of the desired
space ${\B^2_1}$, namely
$$\B^2_1 = \overline {\B^2_1}- P \overline {\B^2_1}. \eqn\withit$$
We verify this by taking now the boundary, and using
$P\partial = \partial P$, together with \djeri, to find 
$$\eqalign{
\partial \B^2_1 &= (1-P) \,\partial \overline {\B^2_1} \,, \cr
&= \T^2_1 - \I\B^1_2  + \bigl( \Sigma (s, 0) - P \Sigma (s, 0)\,\bigr)\,,\cr
&= \T^2_1 - \I\B^1_2 \,,\cr   } \eqn\ghty$$
using the symmetry of the space $\Sigma (s,0)$.
Having shown that the $\B^2_1$ space defined by \djeri\ and \withit\
has the correct boundary this completes our analysis of the first 
consistency condition.

\section{Iterative construction of the $\B^2_k$ spaces}

The second recursion relation listed in \frsttw\ is an example of the general 
situation which takes the form 
$$\partial \B^2_k =   \K \,\B^1_{k} -  \Bigl( \I \B^1_{k+1} + \cdots \Bigr)
\, . \eqn\gnrl$$
By the consistency condition, the space $\K \B^1_k$, to be called
$\C$,  and the second space,
indicated by parenthesis and to be called $\C'$, 
are two $\B^2$ spaces having
identical boundaries (the specific way of splitting the right hand side
of the recursion relation will not be relevant). 
Our problem is to construct an interpolation between
the  $\B^2$ spaces $\C$ and $\C'$, such that the boundary of the interpolation 
gives the two  spaces in
question. Note that the surfaces involved are all spheres with $k+2$ punctures,
and the spaces $\C$ and $\C'$ have real dimension one higher than
that of the moduli space $\M_{k+2}$ of punctured Riemann spheres without
decoration. Thus each of the two 
$\B^2$  spaces can be thought as built from 
one dimensional fibers over $\M_{k+2}$. 

The idea now is to project $\C$,
which sits in $\wh\P_{k+2}$ (the moduli space of $(k+2)$-punctured
spheres having local coordinates at the punctures) down to
the base $\M_{k+2}$ via the projection $\pi : \wh\P_{k+2}\to \M_{k+2}$. We
then to examine for each  point $p \in \pi (\,\C\, )$ 
the intersection $\pi^{-1} (p) \cap \C\equiv \C_p$, namely 
the intersection of the (infinite dimensional) fiber over $p$ with the
space $ \C$. This intersection $\C_p$ must be a 
one dimensional subspace of $\C$, since the dimensionality of $\C$ 
exceeds that of $\M_{k+2}$ by one. We choose some consistent parametrization
of the surfaces on each of the spaces $\C_p$; namely, they are parametrized
by a $t\in [0,1]$ such that for nearby points $p,p'$ and nearby parameters
$t,t'$ we get nearby surfaces, and the parametrization is compatible
with the antisymmetry properties on the exchange of special punctures.
The space $\C$ can be reconstructed as the union over all $p$
of the spaces $\C_p$, namely $\C = \cup_p\C_p$.

The one dimensional space $\C_p$ must have two endpoints on 
${\p\C}~$.\foot{It is possible, in principle, that $\C_p$ takes the form
of a disjoint union of segments, each with endpoints on $\partial \C$. In 
this case one would use several parameters, all running from zero to one.} 
The  space $ \C'_p \equiv \pi^{-1} (p) \cap \C' $
 must also have endpoints on $\partial\C'_p$, and the endpoints of 
$\C_p$ and $\C'_p$ must coincide given 
that $\partial (\C- \C')=0$, and $\partial (\C- \C')=\cup_p  (\partial\C_p-
\partial\C_p')$.  We now construct a space ${\cal D}_p$ interpolating
 between the parametrized spaces
$\C_p$ and $\C'_p$ by canonical interpolation on the local coordinates 
at the punctures
for each value of the parameter. This interpolation
is all that needs to be done
since all the surfaces agree as punctured Riemann surfaces without
decoration. Note that by construction the boundary of
${\cal D}_p$ is simply $\C_p - \C'_p$. The space ${\cal D}_p$ is
built for all points
$p$, and we take  $\B^2_k = \cup_p{\cal D}_p$. It follows that 
$\partial \B^2_k = \cup_p \partial{\cal D}_p = \cup_p (\C_p - \C'_p)=
\C -\C'$, as desired. This completes the construction.

\ack 
I wish to thank R. Cohen for preparing the figures.


\refout

\bye

The second recursion relation listed in \frsttw\ is an example of the general 
situation which takes the form 
$$\partial \B^2_k =   \K \,\B^1_{k} -  \Bigl( \I \B^1_{k+1} + \cdots \Bigr)
\, . \eqn\gnrl$$
By the consistency condition, the space $\K \B^1_k$, to be called
$\C$  and the second space,
indicated by parenthesis, and to be called $\C'$ 
are two $\B^2$ spaces having
identical boundaries (this specific way of splitting the right hand side
of the recursion relation will not be relevant to the analysis). 
Our problem is to construct an interpolation between
the  $\B^2$ spaces $\C$ and $\C'$, such that the boundary of the interpolation 
gives the two  spaces in
question. Note that the surfaces involved are all spheres with $k+2$ punctures,
and all the spaces in question have real dimension one higher than that of
the moduli space $\M_{k+2}$ of punctured Riemann spheres  (without
local coordinates at the punctures). Thus each of the two 
$\B^2$  spaces can be thought as built from 
one dimensional fibers over $\M_{k+2}$. We will construct an
interpolation between the primitive domains
$\overline\C$ and $\overline\C'$ and then we will 
extend it to the complete spaces.

A primitive domain $\overline \C$ of a $\B^2$ space $\C$ will have a boundary 
 $ \partial\hskip2pt \overline\C$ composed
of $\overline {\partial \C}$, a primitive domain of the boundary of $\C$,
 plus boundaries to be called symmetry planes, where surfaces
are symmetric under the exchange of 
at least two of the special punctures. The idea now is to project $\ov\C$,
which sits in $\wh\P_{k+2}$ (moduli space with data of 
local coordinates at punctures) down to
the base $\M_{k+2}$ via the projection $\pi$, and then to examine for each 
point $p \in \pi (\,\ov\C\, )$ 
the intersection $\pi^{-1} (p) \cap \ov \C\equiv \C_p$, namely 
the intersection of the (infinite dimensional) fiber over $p$ with the
space $\ov \C$. This intersection $\C_p$ must be a 
one dimensional subspace of $\ov\C$, since the dimensionality of $\ov\C$ 
exceeds that of $\M_{k+2}$ by one. We choose some consistent parametrization
of the surfaces on each of the spaces $\C_p$, namely they are parametrized
by a $t\in [0,1]$ such that for nearby points $p,p'$ and nearby parameters
$t,t'$ we get nearby surfaces. The one dimensional space 
$\pi^{-1} (p) \cap \ov \C\equiv \C_p$ must be of either of the following types:

\noindent
(i) it has two endpoints on $\ov{\p\C}$, or

\noindent
(ii) it has one endpoint on $\ov{\p\C}$ and one endpoint on a symmetry
plane, or  

\noindent
(iii) it has two endpoints on symmetry planes, or

\noindent
(iv) it has no boundary.

\noindent
In case (i) the other space $\pi^{-1} (p) \cap \ov \C' \equiv \C'_p$
 must have the same structure, and the endpoints must coincide, as they
represent true boundaries of the spaces and must therefore cancel. 
In this case we must simply connect the parametrized spaces
$\C_p$ and $\C'_p$ by an interpolation on the local coordinates at the punctures
for each value of the parameter. 
Case (ii) is the possibility that took place
in our construction of $\B^2_1$ in the previous section. In this case
$\C'_p$ must have an endpoint coinciding with that endpoint of $\C_p$
on $\overline {\partial \C}$, and an endpoint on a symmetry plane.
We have parametrized spaces ${\C'}_p$ and $\C_p$
that again can be connected by coordinate interpolation.

\noindent
Cases (iii) and (iv) are of a different type. If $\C_p$ is either of
type (iii) or (iv), then also ${\C'_p}$ must be of type (iii) or (iv),
in any order. This is the case because the boundaries do not show up
as true boundaries.  In either case there is no matching to be done,
and one treats $\C_p$ and $\C'_p$ independently, following the same rules
in both cases. If $\C_p$ is of type (iii) we take the parameter $t$ to have
values zero and one at the endpoints and construct interpolations between
the surfaces at $t$ and $1-t$, for all $t\in [0,\half]$. This defines
a space whose boundary will be $\C_p$ plus the symmetric interpolation
between its endpoints. If $\C_p$ is of type (iv)
it is clear that $\C_p$ is the boundary of a disk sitting completely in
the fiber over $p$, the disk can be determined canonically by again 
interpolating between $t$ and $1-t$.

The above cases define the 
interpolation ${\cal D}$ between  $\overline \C$ and $\overline{\C'}$. 
It accounts for the desired boundaries, but  
has additional boundaries on symmetry
planes.  The total
 interpolation for the $\B^2$ spaces $\C$ and $\C'$ is given
by $(1-P) {\cal D}$. In this interpolation the boundaries lying
on symmetry planes cancel out in pair. This completes our construction.

{\bf alternative proof}

\bye